\documentclass{article}%
\usepackage{amsfonts}
\usepackage{amssymb}
\usepackage{amsmath}
\usepackage{graphicx}%
\providecommand{\U}[1]{\protect\rule{.1in}{.1in}}
\begin{document}

\title{The Mathematics of Income Property Valuation}
\author{David Ellerman\\Faculty of Social Sciences\\University of Ljubljana, Slovenia\\david@ellerman.org\\orcid.org/0000-0002-5718-618X}
\maketitle

\begin{abstract}
\noindent This paper surveys and generalizes the main valuation formulas used
in real estate valuation and presents unified proofs. The results are
otherwise scattered in obscure journals and books while the proofs are rarely
available to researchers in the field. The material was originally developed
so that it could be used by mathematically-trained appraisers and researchers
in the former Soviet Union and in other transition economies that were
starting their real estate valuation profession.



\end{abstract}

%\tableofcontents

\section{Introduction}

To what extent is real estate or income property valuation a science? Today, a
property appraiser often plugs in measurements and other data into a computer
program which determines the \textquotedblleft comparables\textquotedblright%
\ in a developed market, and then a database is consulted to see the recent
prices of comparable properties in those markets. But what ultimately gives
any market-based objectivity to those comparables? Is the market in a
temporary bubble or depressed state? What about unique real estate properties
with no comparables? What about the examples like the post-communist countries
where there were no prior real estate markets? 

The underlying basis for market-based values is the treatment of a real estate
property as an \textit{income property} \cite{friedman:reapp}. This involves
the calculation of a value based on present and future costs and revenues
discounted back into one net present value. The topic of this paper is the
exposition and mathematical proofs of the basic and advanced formulas that
could be used to compute the net present value of an income property. 

There are a number of reasonably complex formulas that are used in the income
approach to real estate appraisal, particularly as developed in the United
States. The necessary assumptions and the proofs of these formulas are usually
to be found only in a few scarce journal articles in the United States or in
out-of-print books. Hence we have attempted to give here, all in one place,
fresh algebraic derivations of the major formulas to make them available to
technically adept students and practitioners.

A number of new results are also presented:

(1) a general formula for the valuation of changing income streams defined by
linear recurrence relations which has all the usual formulas for valuing
changing income streams as special cases (e.g., straight line changing
annuity, constant ratio changing annuity, and Ellwood $J$ premise),

(2) an analysis of the straight line and Hoskold capitalization methods which
shows that both methods are appropriate for certain declining income streams
where the income decline can be motivated as the interest losses resulting
from a hypothetical capital recovery sinking fund using a substandard rate
(below the discount rate), and

(3) a general theorem about amortization tables where the principal reductions
can be arbitrarily specified and an application of the theorem to give an
alternative proof of the main result about the Hoskold capitalization method.

\section{The Six Functions of One}

\subsection{The Amount of One at Compound Interest}

Throughout our discussion, we will assume that future amounts of money can be
discounted back to present values or that present amounts can be compounded
into future values using a discount rate $r$ (or $i$) per period. The periods
could be years, months, or any other fixed time period. Unless otherwise
stated, the formulas will always assume that the interest rate (\% per period)
and the units of time are stated using the same period of time. The discount
rate may be taken as including the risk-free interest rate and a consideration
for risk and illiquidity.

The first basic formula

\begin{figure}[h]
\centering
\includegraphics[width=0.7\linewidth]{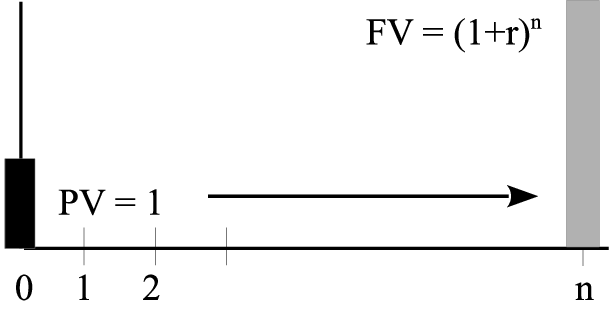}  \caption{Amount of one at
compound interest}%
\label{fig:fv-pix}
\end{figure}

\begin{center}
$FV=PV(1+r)^{n}$
\end{center}

\noindent states that given the present value of $PV$, that is equivalent on
the market to the future value after n periods of $FV=PV(1+r)^{n}$. If $PV=1$,
then we have the \textit{amount of one at compound interest} given in the
tables. The present value is said to be "compounded" into the future value as
illustrate in Figure \ref{fig:fv-pix}.

\subsection{The Present Value Reversion of One}

For each basic function of one, the inverse or reciprocal is also a function
of one. The inverse of the amount of one at compound interest is the present
value reversion of one.

\begin{center}
$PV=\frac{FV}{\left(  1+r\right)  ^{n}}$.
\end{center}

\begin{figure}[h]
\centering
\includegraphics[width=0.7\linewidth]{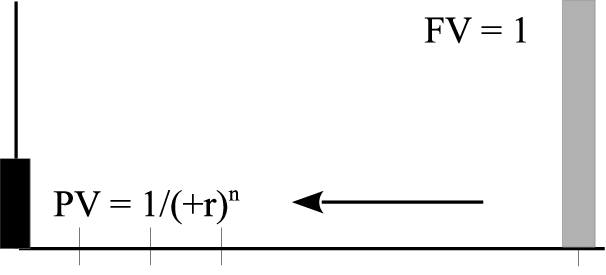}  \caption{Discounted present
value of one}%
\label{fig:pv-pix}%
\end{figure}

Given a future amount $FV$ at the end of the $n^{th}$ period, the equivalent
present value (at time zero) is obtained by dividing by the factor of
$(1+r)^{n}$ as show in Figure \ref{fig:pv-pix}.

The future value is said to be \textquotedblleft discounted\textquotedblright%
\ to the present value.

\subsection{The Present Value of an Ordinary Annuity of One}

Suppose we want to pay off a loan with a series of equal payments at the times
$t=1,2,...,n$ (i.e., at the end of the first period and the end of each other
period up to and including the $n^{th}$ period). We consider a series of equal
payments of one. Each payment is discounted back to a present value using the
present-value-of-one formula (taking care to use the right time period). Since
the results are all amounts of money at the same time, they can be
meaningfully added together to get the total present value of the series of
equal payments. It is called the \textit{present value of an ordinary annuity
of one} and will be denoted $a(n,r)$.

\begin{center}
$a\left(  n,r\right)  =\frac{1}{\left(  1+r\right)  ^{1}}+\frac{1}{\left(
1+r\right)  ^{2}}+...+\frac{1}{\left(  1+r\right)  ^{n}}=\sum_{k=1}^{n}%
\frac{1}{\left(  1+r\right)  ^{k}}=\frac{1-\frac{1}{\left(  1+r\right)  ^{n}}%
}{r}$.
\end{center}

\begin{figure}[h]
\centering
\includegraphics[width=0.7\linewidth]{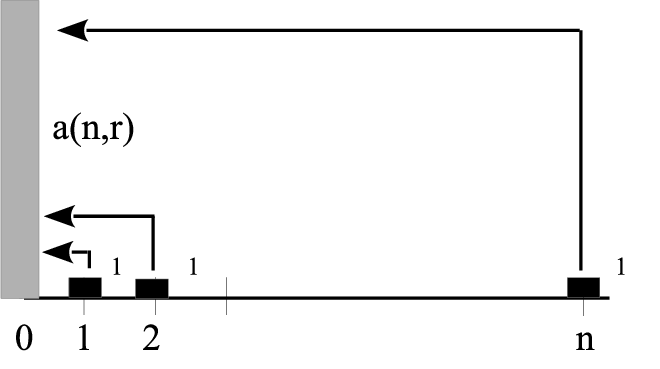}  \caption{Present value of
an ordinary annuity of one}%
\label{fig:ann-pix}%
\end{figure}

Given a series of equal payments $PMT$ at $t=1,2,...,n$, their present value
is $PMTa(n,r)$. Those payments would pay off a loan at time zero of that
principal value of $PV=PMTa(n,r)$ as shown in Figure \ref{fig:ann-pix}.

\subsection{The Installment to Amortize One}

If we are given the equal payments $PMT$, we can use the present value of an
annuity of one $a(n,r)$ to calculate the corresponding principal value
$PV=PMTa(n,r)$. But if we are given the principal $PV$ for a loan, then we can
use the reciprocal $1/a(n,r)$ to calculate the equal installment payments
$PMT=PV/a(n,r)$ that would pay off the loan. The equal installment payments
are said to \textquotedblleft amortize\textquotedblright\ the loan. If the
loan was for $PV=1$, then the reciprocal amount $PMT=1/a(n,r)$ is called
\textit{the installment to amortize one}.

\begin{center}
$PMT=\frac{1}{a\left(  n,r\right)  }=\frac{1}{\frac{1}{\left(  1+r\right)
^{1}}+\frac{1}{\left(  1+r\right)  ^{2}}+...+\frac{1}{\left(  1+r\right)
^{n}}}=\frac{r}{1-\frac{1}{\left(  1+r\right)  ^{n}}}$
\end{center}

\begin{figure}[h]
\centering
\includegraphics[width=0.7\linewidth]{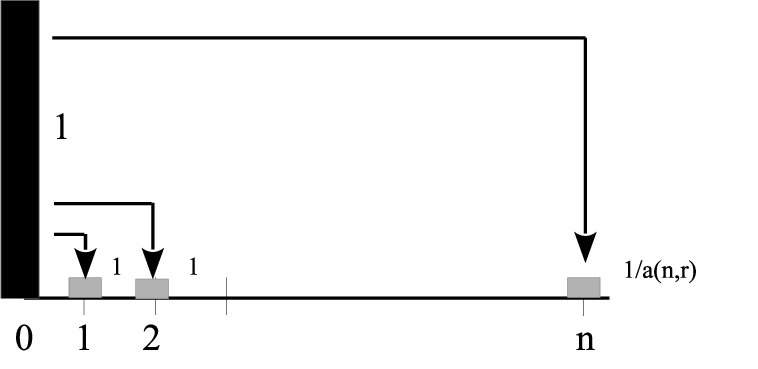}  \caption{The installment
payments to amortize loan of one}%
\label{fig:amt-pix}%
\end{figure}

We can think of the present value $PV=1$ as \textquotedblleft
growing\textquotedblright\ into the equal series of $1/a(n,r)$ amounts.
Suppose the present amount of one is deposited in a bank account being the
compound interest rate of $r$ per period. At the end of period $1$, the amount
$1/a(n,r)$ can be withdrawn from the account leaving the remainder to
accumulate interest. In a similar manner, the amount $1/a(n,r)$ can be
withdrawn at the end of period $2$ and so forth through period $n$. The last
withdrawal of $1/a(n,r)$ at time $n$ would reduce the bank account balance to
zero as illustrated in Figure \ref{fig:amt-pix}.

\subsection{The Accumulation of One per Period}

Suppose that instead of considering the present value of a series of equal
payments, we consider the future value at time $n$ of a series of equal
amounts at time $1,2,...,n$. This practice of depositing equal amounts over a
series of time periods and letting them accumulate to a future amount is
called a \textit{sinking fund}. Each deposit in the fund can be compounded to
a future value at time $n$ and the future values can be added together to get
the total accumulated value of the sinking fund. If each deposit is one, then
the total future amount is called \textit{the accumulation of one per period}
and is denoted $s(n,r)$.

\begin{center}
$s(n,r)=\left(  1+r\right)  ^{n-1}+\left(  1+r\right)  ^{n-2}+...+\left(
1+r\right)  ^{1}+1=\sum_{k=0}^{n-1}\left(  1+r\right)  ^{k}=\frac{\left(
1+r\right)  ^{n}-1}{r}$
\end{center}

Since this accumulation of one per period just restates the present value of
an annuity of one as a future value at time $n$, we have

\begin{center}
$s\left(  n,r\right)  =\left(  1+r\right)  ^{n}a\left(  n,r\right)  $. See
Figure \ref{fig:accm-pix}.

\begin{figure}[h]
\centering
\includegraphics[width=0.7\linewidth]{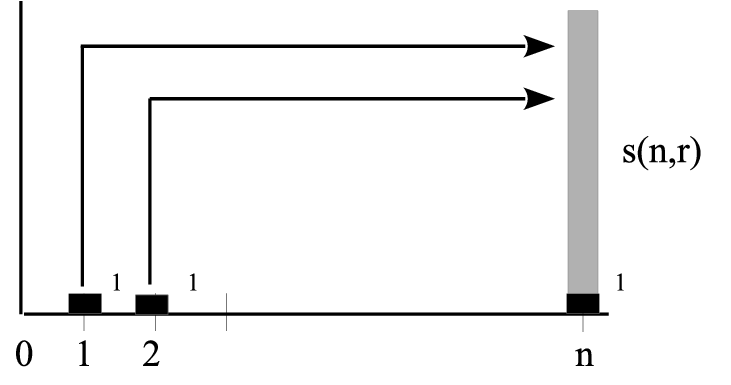}  \caption{The accumulation
of one per period}%
\label{fig:accm-pix}%
\end{figure}
\end{center}

\subsection{The Sinking Fund Factor}

For the inverse function, we know the desired value of the accumulated fund
$FV$ at time $n$ and we compute the sinking fund deposit (or payment $PMT$
into the fund) at times $1,2,...,n$ that would accumulate to the desired
amount $FV$. That deposit is called \textit{the sinking fund factor} and will
be denoted $SFF$.

\begin{center}
$SFF(n,r)=\frac{1}{s\left(  n,r\right)  }=\frac{1}{\left(  1+r\right)
^{n-1}+\left(  1+r\right)  ^{n-2}+...+\left(  1+r\right)  ^{1}+1}=\frac
{r}{\left(  1+r\right)  ^{n}-1}$
\end{center}

\begin{figure}[h]
\centering
\includegraphics[width=0.7\linewidth]{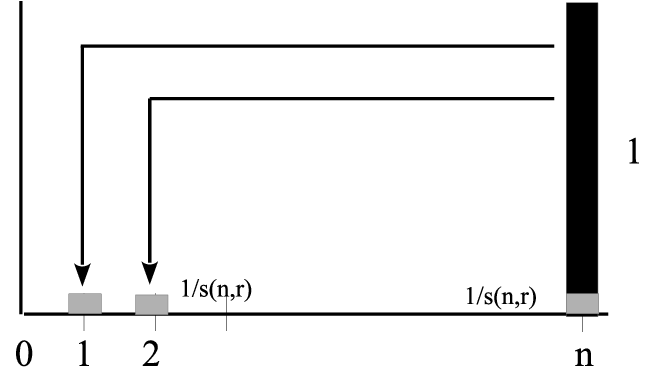}  \caption{The sinking fund
factor}%
\label{fig:sff-pix}%
\end{figure}

The sinking fund factor $SFF$ \textquotedblleft discounts\textquotedblright%
\ the future value of the fund $FV=1$ back into a series of equal amounts. If
you had the promise to receive the future value of one at time $n$, then it
would be equivalent for you to receive the series of equal payments
$SFF=1/s(n,r)$ at the times $1,2,...,n$ as in Figure \ref{fig:sff-pix}.

\subsection{Summary of Six Functions}

The six functions can be divided into two groups: three functions and their
inverses \cite{friedman:reapp} as in the following Table 1.

\begin{center}%
\begin{tabular}
[c]{|c|c|}\hline
Function & Inverse Function\\\hline\hline
Amount of One at Compound Interest & Present Value Reversion of One\\
$\left(  1+r\right)  ^{n}$ & $\left(  1+r\right)  ^{-1}$\\\hline\hline
Present value of an Annuity of One & Installment to Amortize One\\
$a\left(  n,r\right)  =\frac{1}{\left(  1+r\right)  ^{1}}+\frac{1}{\left(
1+r\right)  ^{2}}+...+\frac{1}{\left(  1+r\right)  ^{n}}$ & $\frac{1}{a\left(
n,r\right)  }=\frac{1}{\frac{1}{\left(  1+r\right)  ^{1}}+\frac{1}{\left(
1+r\right)  ^{2}}+...+\frac{1}{\left(  1+r\right)  ^{n}}}$\\\hline\hline
Accumulation of One per Period & Sinking Fund Factor\\
$s(n,r)=\left(  1+r\right)  ^{n-1}+\left(  1+r\right)  ^{n-2}+...+\left(
1+r\right)  ^{1}+1$ & $\frac{1}{s\left(  n,r\right)  }=\frac{1}{\left(
1+r\right)  ^{n-1}+\left(  1+r\right)  ^{n-2}+...+\left(  1+r\right)  ^{1}+1}%
$\\\hline
\end{tabular}

Table 1: Three basic functions and their inverses.
\end{center}

\subsection{Amortization Tables}

Let us now consider a loan with the principal of $PV$ which is to be paid off
with equal payments $PMT=PV/a(n,i)$ at times $1,2,...,n$. Each payment $PMT$
will pay some interest and pay some principal. The interest payments just
service the loan; they do not reduce the principal balance. Only the remaining
part of $PMT$ can be considered as a principal payment or principal reduction.
How much of each payment is considered as interest payment and how much as
principal payment? The conventional way to compute interest and principal
portions of loan payments is to assume that all the interest due at any time
is taken out of the payment, and the remainder of the payment is principal reduction.

Let $Bal(k)$ be the principal balance due on the loan after the payment is
made at the end of the $k^{th}$ period. The loan begins with $Bal(0)=PV$. At
the end of the first period, the interest due is $iPV=iBal(0)$. Subtracting
from the payment $PMT$ gives the principal portion of the payment
$PMT-iBal(0)$. The new balance is the old balance reduced by the principal
payment: $Bal(1)=Bal(0)-(PMT-iBal(0))=\left(  1+i\right)  Bal\left(  0\right)
-PMT$. In general, the interest due at the end of the $k^{th}$ period is
$iBal(k-1)$ so the principal reduction by the $k^{th}$ payment is:

\begin{center}
$PR(k)=Bal(k-1)-Bal\left(  k\right)  =PMT-iBal(k-1)$.
\end{center}

The new balance at the end of the $k^{th}$ period is:

\begin{center}
$Bal(k)=Bal(k-1)-PR(k)=\left(  1+i\right)  Bal(k-1)-PMT$.
\end{center}

Thus we also have for $k=2,...,n$:

\begin{center}
$PR\left(  k\right)  =PMT-iBal\left(  k-1\right)  =PMT-i\left(  Bal\left(
k-2\right)  -PR\left(  k-1\right)  \right)  $

$=PMT-i\left(  Bal\left(  k-2\right)  -\left[  PMT-iBal\left(  k-2\right)
\right]  \right)  $

$=\left(  1+i\right)  PMT-\left(  1+i\right)  i\left(  Bal\left(  k-2\right)
\right)  =\left(  1+i\right)  PR\left(  k-1\right)  $
\end{center}

\noindent so the Principal Reductions increase each period by $\left(
1+i\right)  $.

The final payment at time $n$ pays off the remaining balance of the loan so
$PR(n)=Bal(n-1)$ and $Bal(n)=0$.

The computation of these interest and principal portions is usually presented
in an amortization table such as Table 2.

\begin{center}%
\begin{tabular}
[c]{|c|c|c|c|c|c|}\hline
Period & Beg. Balance & Payment & Interest & Prin. Reduction & End
Bal.\\\hline\hline
$1$ & $PV$ & $PMT$ & $iPV$ & $PMT-iPV$ & $Bal(1)$\\\hline
$2$ & $Bal\left(  1\right)  $ & $PMT$ & $iBal\left(  1\right)  $ &
$PMT-iBal\left(  1\right)  $ & $Bal\left(  2\right)  $\\\hline
$\cdots$ & $\cdots$ & $\cdots$ & $\cdots$ & $\cdots$ & $\cdots$\\\hline
$n-1$ & $Bal\left(  n-2\right)  $ & $PMT$ & $iBal\left(  n-2\right)  $ &
$PMT-iBal\left(  n-2\right)  $ & $Bal\left(  n-1\right)  $\\\hline
$n$ & $Bal\left(  n-1\right)  $ & $PMT$ & $iBal\left(  n-1\right)  $ &
$PMT-iBal\left(  n-1\right)  $ & $Bal\left(  n\right)  =0$\\\hline
\end{tabular}

Table 2: Amortization table
\end{center}

\subsection{Some Formulas of Financial Mathematics}

To derive a formula for $Bal(k)$ the balance due at the end of the $k^{th}$
period for a loan of principal $PV$, we first derive the formula for $bal(k)$,
the balance due at time $k$ for a loan of principal $1$. Then we will have:
$Bal(k)=PVbal(k)$.

We know that $a(n,i)$ is the present value of payments of $1$ at the end of
each period $1,...,n$. This sum can be divided into two parts, the present
value of the first $k$ payments which is $a(k,i)$, and the value of last $n-k$
payments at time $k$, namely $a(n-k,i)$ discounted backed to time $0$ by
dividing by $(1+i)^{k}$:

\begin{center}
$a\left(  n,i\right)  =a\left(  k,i\right)  +\frac{a\left(  n-k,i\right)
}{\left(  1+i\right)  ^{k}}$.
\end{center}

Multiplying both sides by $(1+i)^{k}/a(n,i)$:

\begin{center}
$(1+i)^{k}=\frac{(1+i)^{k}a\left(  k,i\right)  }{a\left(  n,i\right)  }%
+\frac{a\left(  n-k,i\right)  }{a\left(  n,i\right)  }$
\end{center}

\noindent and rearranging yields the formula for $bal(k)$:

\begin{center}
$bal\left(  k\right)  =\frac{a\left(  n-k,i\right)  }{a\left(  n,i\right)
}=(1+i)^{k}\left[  1-\frac{a\left(  k,i\right)  }{a\left(  n,i\right)
}\right]  $.
\end{center}

The reasoning is that if the principal of the loan is $1$, then each payment
is $1/a(n,i)$. The balance at time $k$, $bal(k)$, is the present value at that
time of the last $n-k$ payments so we have the above formula.

We will later have occasion to use the Portion Paid $PP(k)$ of a loan at time
$k$ which is simply one minus the balance of the loan of one at that time:

\begin{center}
$PP\left(  k\right)  =1-bal\left(  k\right)  =1-\frac{a\left(  n-k,i\right)
}{a\left(  n,i\right)  }=1-(1+i)^{k}\left[  1-\frac{a\left(  k,i\right)
}{a\left(  n,i\right)  }\right]  $.
\end{center}

\begin{figure}[h]
\centering
\includegraphics[width=0.7\linewidth]{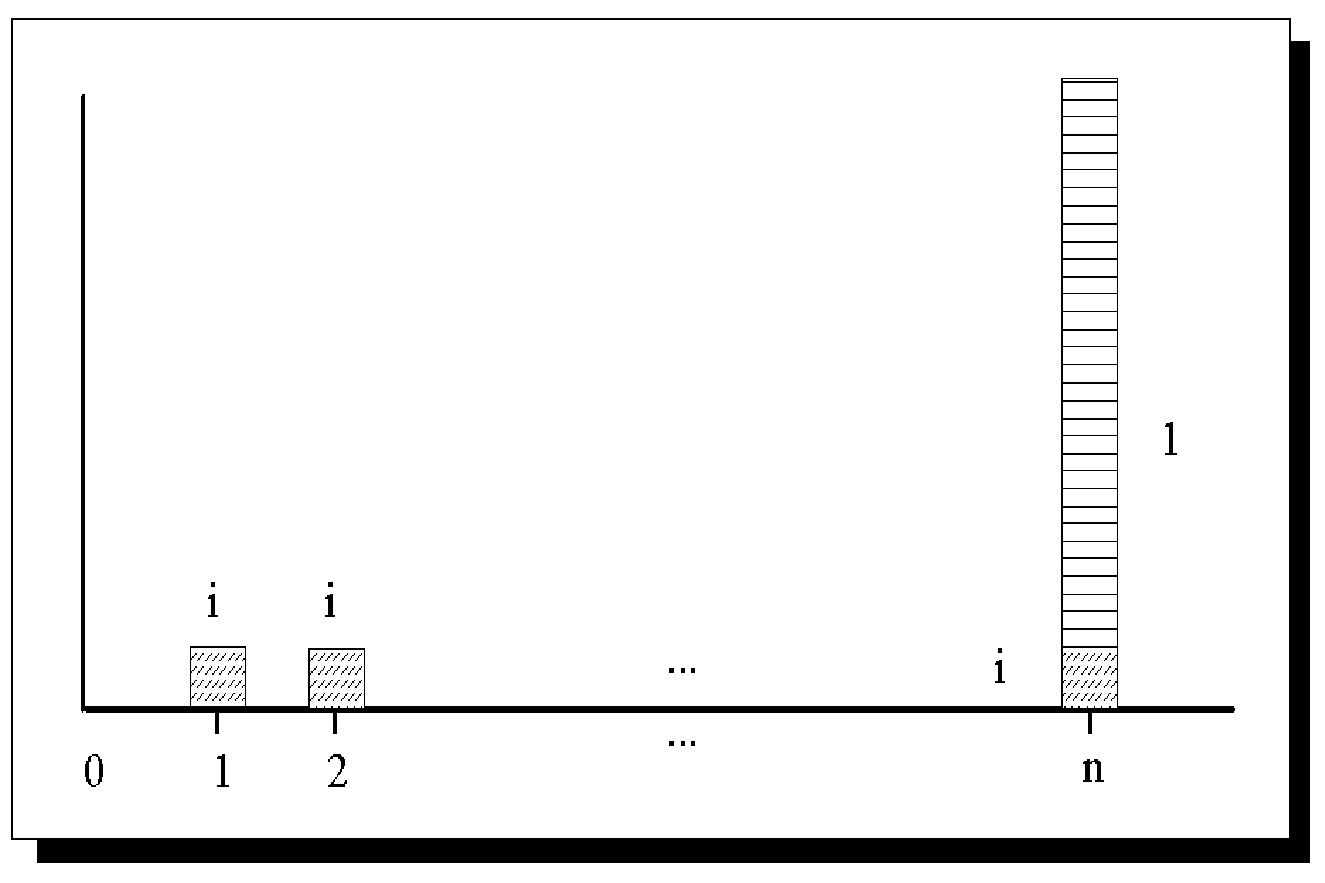}
\caption{Interest only payments plus balloon payment}%
\label{fig:interest-only-baloon}%
\end{figure}

We have seen that for the case of $PV=1$, the $n$ payments $PMT=1/a(n,i)$ will
pay off the loan. That is, the present value of those equal payments is the
principal amount $1$ of the loan. But there are many other future series of
payments--unequal payments--which would also have that present value. For
instance, we could pay the same interest on one of $i$ at the end of each
period and pay no principal until the end of the $n^{th}$ period when we pay
all the principal in one "balloon payment" of one as in Figure
\ref{fig:interest-only-baloon}.

\begin{figure}[h]
\centering
\includegraphics[width=0.7\linewidth]{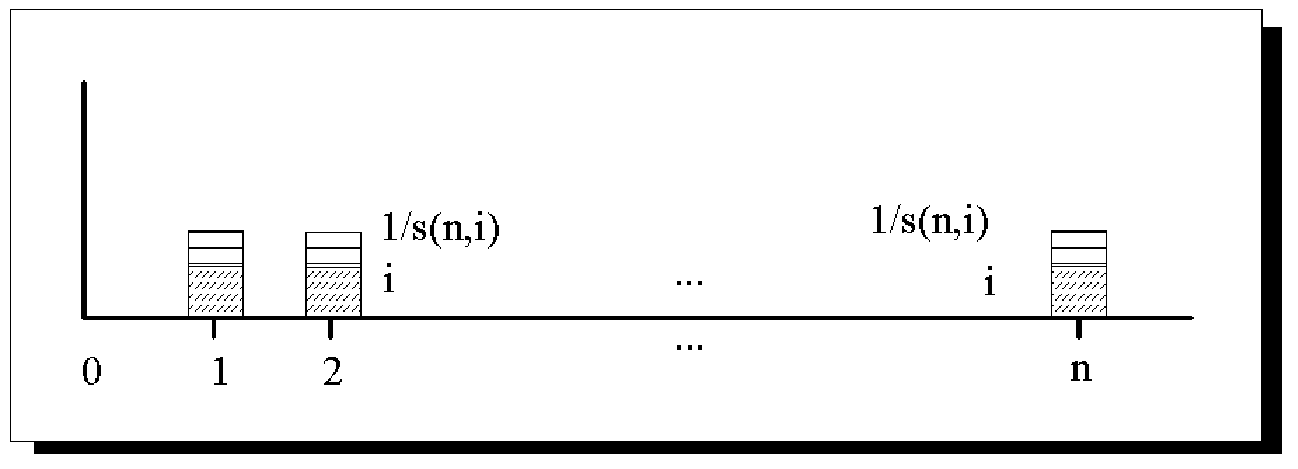}  \caption{Sinking fund
factor SFF}%
\label{fig:1sni}%
\end{figure}

That unequal series of payments also has the present value of one. But how
will we make the balloon payment? Suppose we make a sinking fund deposit of
$SFF=1/s(n,i)$ at times $1,2,...,n$. Those deposits will accumulate to $1$ at
time $n$ to give precisely the balloon payment. But that means that the equal
payments at times $1,2,...,n$ of the interest $i$ plus the sinking fund factor
will also have the present value of one (since that pays off that loan) as
shown in Figure \ref{fig:1sni}.

But we have another series of equal payments at $t=1,2,...,n$ with the present
value of one, namely the installments of amortize one $1/a(n,i)$. Hence the
two payments must be equal, and we have the important formula:

\begin{center}
$\frac{1}{a\left(  n.i\right)  }=\frac{1}{s\left(  n,i\right)  }+i$.
\end{center}

In words, the installment to amortize one is the sum of sinking fund factor
plus the discount rate.

%\quad

\section{Direct Capitalization Formulas}

\subsection{The IRV Formula}

Another useful formula can be derived by considering an infinite series of
equal payments called a "perpetuity." We know that the present value of a
finite series of $n$ payments $PMT$ at$t=1,2,...,n$ is

\begin{center}
$PV=\frac{PMT}{a\left(  n,i\right)  }=PMT\frac{1-\frac{1}{\left(  1+i\right)
^{n}}}{i}$.
\end{center}

If the series of payments goes on to infinity then we simply take
$n\rightarrow\infty$ in the formula with takes the present value of one
$1/(1+i)^{n}$ to zero. Thus we have the:

\begin{center}
$PV=\frac{PMT}{i}.$

Perpetuity Capitalization Formula
\end{center}

\begin{figure}[h]
\centering
\includegraphics[width=0.5\linewidth]{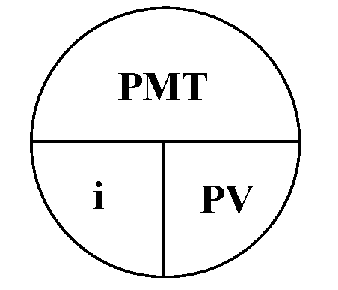}  \caption{Perpetuity
pie diagram}%
\label{fig:pie-diagram}%
\end{figure}

This is a very simple and convenient formula which can be presented in a "pie
diagram" as in Figure \ref{fig:pie-diagram}.

For a perpetuity payment of $PMT$ per period, one can cover up a symbol in the
pie diagram to find the formula for that amount. Cover up $PV$, and you see
the $PV=PMT/i$. Cover up $PMT$, and you see that the perpetual payment with
the present value $PV$ is $PMT=iPV$.

Because of the simplicity of this type of formula, many practitioners would
like to put the more complicated formulas encountered before into the same
format. That is usually possible, and the results are called "direct
capitalization formulas."

Consider, for example, the finite series of payments $PMT$ at $t=1,2,...,n$
with the present value $PV=PMTa(n,i)$. We can rewrite $a(n,i)$ as the
reciprocal of its reciprocal so that the formula is:

\begin{center}
$PV=\frac{PMY}{1/a\left(  n,i\right)  }$.
\end{center}

\begin{figure}[h]
\centering
\includegraphics[width=0.5\linewidth]{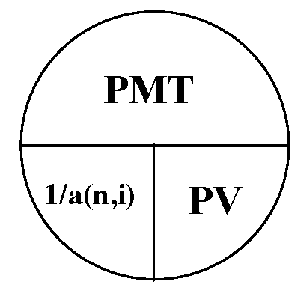}  \caption{Cap rate
pie diagram}%
\label{fig:1-anicap-rate}%
\end{figure}

Thus we see that $1/a(n,i)$ can be thought of as rate used to transform or
"capitalize" the amount of the equal payments $PMT$ into the present value
$PV$. It is then called a "capitalization rate" to distinguish it from the
discount rate $i$ as in Figure \ref{fig:1-anicap-rate}.

\begin{figure}[h]
\centering
\includegraphics[width=0.5\linewidth]{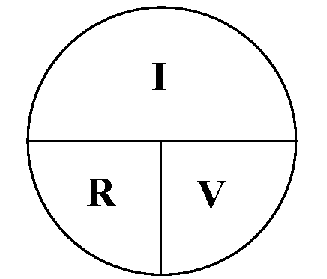}  \caption{$I=RV$ Formula}%
\label{fig:ieqrv}%
\end{figure}

In the real estate valuation literature, the amount $PMT$ is the income $I$
(e.g., the net operating income $NOI$ of an income-producing property), the
capitalization rate is denoted as $R$, and the present value is just called
the value $V$. Thus we have the famous $I=RV$ formula as in Figure
\ref{fig:ieqrv}.

We previously saw that the capitalization rate $R=1/a(n,i)$ could be expressed
as the sum of the sinking fund factor and the discount rate so we have:

\begin{center}
$V=\frac{I}{1/a\left(  n,i\right)  }=\frac{I}{i+1/s\left(  n,i\right)  }$.
\end{center}

Cross-multiplying shows that each income $I$ is the sum of the amount
$V/s(n,i)$ and $iV$. The latter is simply the interest on the value $V$ and it
is called the "return ON investment." Since the amount $V/s(n,i)$ is the
sinking fund deposit which would accumulate to $V$ at time $n$, it is called
the "capital recovery" part of the income or the "return OF investment."

\begin{center}
$I=iV+V/s(n,i)=$ Return on investment + Return of investment.
\end{center}

\subsection{The Cap-Rate Style of Reasoning}

There are various ways to express the formulas of financial mathematics. The
income approach to real estate appraisal, particularly in the USA, has
developed a strong tendency to express formulas in a certain way which, in
turn, promotes a certain "style" of reasoning. In making the following remarks
about this "cap-rate style" our purpose is not to criticize it but only to
point out that it is a free choice and that other choices would also be quite
possible. Let us begin with the basic formula to capitalize a perpetual income
stream of one dollar payments or incomes:

\begin{center}
$V=\frac{1}{i}$.
\end{center}

How should the formula be changed to value a truncated income stream stopping
at time $n$? The "cap-rate style" is to change the formula by modifying the
capitalization rate to account for the truncation of the income stream at
$t=n$ to obtain:

\begin{center}
$a\left(  n,i\right)  =\frac{1}{i+\frac{1}{s\left(  n,i)\right)  }}$.
\end{center}

The new formula is explained using the reasoning about "return on investment"
and "return of investment." Since the income stream terminates, the underlying
asset has wasted away so the capitalization rate must be "loaded" with the
sinking fund factor $SFF(n,i)=1/s(n,i)$ to account for the return of investment.

There is, however, another perfectly equivalent way to modify the perpetuity
formula to account for the truncation of the income stream. Instead of
changing the denominator (the capitalization rate), change the numerator (the
income). Instead of loading the cap rate, we can make a deduction from the
income ($1$ per year) to turn it into a perpetual income stream which can then
be capitalized by the same denominator of $i$. What is the deduction to
perpetualize the income--to replace the truncated stream with a perpetual
stream with the same value? From the first income of $1$ at time $1$, set
aside $1/(1+i)^{n}$ which is equivalent to another $1$ at time $n+1$ (i.e.,
which would accumulate to $1$ at time $n+1$ in a sinking fund). From the
second income of $1$ at $t=2$, set aside another $1/(1+i)^{n}$ which
accumulates to $1$ at time $n+2$, and so forth. By making the $1/(1+i)^{n}$
deduction from each of the $1$'s in the truncated income stream, one generates
another stream of $1$'s at the times $n+1,n+2,...,n+n$. The same deductions
are made from those $1$'s, and so forth. Thus the perpetual version of the
truncated income stream of $n$ $1$'s at times $1,2,...,n$ is $1-1/(1+i)^{n}$
which can then be capitalized by dividing by the interest rate:

\begin{center}
$a\left(  n,i\right)  =\frac{1-\frac{1}{\left(  1+i\right)  ^{n}}}{i}$.
\end{center}

This formula is also in the $IRV$ format but it reflects the opposite "income
style" of reasoning, i.e., modify the income instead of modifying the
capitalization rate. Instead of using cap rate reasoning about loading the cap
rate to account for the return of investment, we can use the familiar
reasoning about charging depreciation against income so that an asset can be
replaced when it wastes away. The amount $1/(1+i)^{n}$ is the depreciation
charge against each "$1$" so that it can be replaced $n$ years later to
perpetuate the income stream.

We will see again and again that formulas are developed in real estate
mathematics so that the changes are made to the cap rates, not the incomes.
That in turn determines the style of reasoning and explanation, e.g., loading
cap rates to recover capital instead of charging depreciation against income
to replace capital. It is not a question of right or wrong. Both the formulas
for $a(n,i)$ are correct and equivalent. Some formulas might be more elegantly
expressed by modifying cap rates, while other formulas will find simpler forms
by changing the income terms. The mathematics of real estate valuation has
mostly chosen the cap-rate road, not the income road. With the increasing use
of electronic computers to value uneven cash flows, the form of the formulas
will become less important but the cap-rate style of reasoning will probably
have a longer lasting influence.

\subsection{Adjusting Capitalization Rates for Appreciation and Depreciation}

We are considering a series of payments or income $I$ that terminates at
$t=n$. There is no further value after that time so this corresponds in real
estate valuation to an asset or property that wastes completely away at $t=n$.
Clearly there are other possibilities so we should see how the formulas in the
capitalization rate format could be adjusted.

For instance, if the asset had the same value $V$ at time $n$ as at time $0$,
then it would be equivalent to the perpetuity of incomes $I$ and the value
would be $V=I/i$. Thus when the asset does not depreciate or appreciate, the
sinking fund factor disappears.

What is the general formula in the capitalization rate format when we have a
series of equal incomes $I$ at $t=1,2,...,n$ and then a future value $FV$
at$t=n$? The total present value would be the usual sum of all the discounted values.

\begin{center}
$V=\frac{I}{\left(  1+i\right)  ^{1}}+\frac{I}{\left(  1+i\right)  ^{2}%
}+...+\frac{I}{\left(  1+i\right)  ^{n}}+\frac{FV}{\left(  1+i\right)  ^{n}%
}=\frac{I}{1/a\left(  n,i\right)  }+\frac{FV}{\left(  1+i\right)  ^{n}}$.
\end{center}

The sinking fund deposits at $t=1,2,...,n$ which accumulate to $FV$ at $t=n$
are $FV/s(n,i)$ and the present value at $t=0$ of those deposits is:

\begin{center}
$\frac{FV}{\left(  1+i\right)  ^{n}}=\frac{FV/s\left(  n,i\right)
}{1/a\left(  n,i\right)  }$.
\end{center}

Substituting in the previous formula yields:

\begin{center}
$V=\frac{I+FV/s\left(  n,i\right)  }{1/a\left(  n,i\right)  }=\frac
{I+FV/s\left(  n,i\right)  }{i+1/s\left(  n,i\right)  }$.
\end{center}

Cross-multiplying and solving for $I$ yields:

\begin{center}
$I=\frac{V-FV}{s\left(  n,i\right)  }+iV=V\left[  \frac{1-FV/V}{s\left(
n,i\right)  }+i\right]  =V\times R^{\ast}$
\end{center}

\noindent where the modified capitalization rate

\begin{center}
$R^{\ast}=\frac{1-FV/V}{s\left(  n,i\right)  }+i$
\end{center}

\noindent reflects the future value $FV$ at $t=n$. When $FV=V$, the
capitalization rate reduces to the discount rate $i$. When $FV=0$, we have the
previous formula $R=1/s(n,i)+i$ where the asset has wasted away at $t=n$.

It is convenient to restate the modified capitalization rate in terms of an
appreciation ratio $\Delta_{0}$ so that $100\Delta_{0}$ is the percentage of
appreciation (and where depreciation would be treated as a negative percent).
The future value is $FV=(1+\Delta_{0})V$. Then the capitalization rate can be
expressed as:

\begin{center}
$R^{\ast}=\frac{1-FV/V}{s\left(  n,i\right)  }+i=\frac{1-\left(  1+\Delta
_{0}\right)  }{s\left(  n,i\right)  }+i=i-\frac{\Delta_{0}}{s\left(
n,i\right)  }+i=i-\Delta_{0}SFF\left(  n,i\right)  $.
\end{center}

\begin{figure}[h]
\centering
\includegraphics[width=0.5\linewidth]{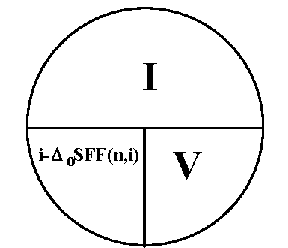}  \caption{Direct
Capitalization Formula with Appreciation or Depreciation}%
\label{fig:loaded-cap-pie}%
\end{figure}

In the real estate literature, the subtraction of the appreciation term to
find the capitalization rate $R^{\ast}$ is called "unloading" for the
appreciation and "loading" for the depreciation (negative appreciation). See
Figure \ref{fig:loaded-cap-pie}.

For no appreciation or depreciation, $\Delta_{0}=0$, and for the fully
depreciated asset, $\Delta_{0}=-1$.

\subsection{Band of Investment Formulas}

We have an income property which yields the net operating income $NOI$ at the
end of each year. A portion $M$ of the value $V$ is financed by a mortgage at
the interest rate $i$ ($M$ is also called "loan to value ratio") so $MV$ is
the principal of the mortgage (note the $MN$ is a product of $M$ times $V$,
not a symbol \textquotedblleft$MV$\textquotedblright). After subtracting the
debt service from the $NOI$, the remainder is the cash return to the equity
holder which is to be discounted at the equity yield rate of $Y$.

A "band of investment" formula is a way to derive a direct capitalization rate
$R$ so that the value $V$ is obtained by capitalizing the $NOI$, i.e.,
$V=NOI/R$. We will derive the formulas for $R$ under a range of assumptions.

In all cases, the value of the property $V$ is the sum of the value of the
equity interest in the property plus the face value of the mortgage:

\begin{center}
Value = Equity + Mortgage Value.
\end{center}

\subsection{Interest-only Loan, No Change in Asset Value, and No Sale of
Asset}

It is assumed that the asset yields an infinite stream of annual net operating
incomes $NOI$ and that the mortgage is an interest-only loan so the debt
service is $MVi$. Thus the equity stream capitalizes to the value $[NOI-$
$MVi]/Y$ and the mortgage value is $MV$ so the total value equation is:

\begin{center}
$V=\frac{NOI-MVi}{Y}+$ $MV$.
\end{center}

Collecting the $V$-terms to the left side we have:

\begin{center}
$V\left[  1+\frac{Mi}{Y}-M\right]  =\frac{NOI}{Y}$
\end{center}

\noindent so dividing and rearranging yields:

\begin{center}
$V=\frac{NOI}{Y\left[  1+\frac{Mi}{Y}-M\right]  }=\frac{NOI}{\left[
Mi+\left(  1-M\right)  Y\right]  }$
\end{center}

Thus the value $V$ could be obtained by capitalizing the $NOI$ at the direct
capitalization rate $R$ where:

\begin{center}
$R=Mi+\left(  1-M\right)  Y$
\end{center}

\noindent is the weighted average of the interest rate $i$ and the equity
yield rate $Y$ with the weights being the mortgage and equity portions of the value.

\subsection{Interest-only Loan, No Change in Asset Value, and Resale of Asset
after H Years.}

The conditions are as above except that the asset is sold for the value $V$
(no change in asset value) after the holding period of $H$ years. Then the
value of the equity is the present value of equity cash return over the
holding period plus the present value of the sales proceeds net of paying off
the mortgage:

\begin{center}
$Equity=a\left(  H,Y\right)  \left[  NOI-MVi\right]  +\left(  1+Y\right)
^{-H}\left[  V-MV\right]  $.
\end{center}

Adding in the mortgage face value yields the value equation:

\begin{center}
$V=a\left(  H,Y\right)  \left[  NOI-MVi\right]  +\left(  1+Y\right)
^{-H}\left[  V-MV\right]  +$ $MV$.
\end{center}

Collecting the $V$-terms to the left yields:

\begin{center}
$V\left[  1+a\left(  H,Y\right)  Mi-\left(  1+Y\right)  ^{-H}\left(
1-M\right)  -M\right]  =a\left(  H,Y\right)  NOI$.
\end{center}

Solving for $V$ and rearranging yields:

\begin{center}
$V=\frac{NOI}{\left[  \frac{\left(  1-M\right)  }{a\left(  H,Y\right)
}+Mi-\frac{\left(  1+Y\right)  ^{-H}\left(  1-M\right)  }{a\left(  H,Y\right)
}\right]  }$
\end{center}

\noindent where the denominator can be written as:

\begin{center}
$R=Mi+\frac{\left(  1-M\right)  \left[  1-\left(  1+Y\right)  ^{-H}\right]
}{a\left(  H,Y\right)  }$.
\end{center}

But $a(H,Y)=[1-(1+Y)^{-H}]/Y$ so we have the previous formula for $R$:

\begin{center}
$R=Mi+\left(  1-M\right)  Y$.
\end{center}

\subsection{Mortgage Amortization over Holding Period, Asset Depreciation
Equal to Mortgage, and Asset Resale after H Years}

Assume that the mortgage with an annual interest rate of $i$ is amortized over
the holding period of $H$ years in $12H$ monthly payments. The monthly payment
for a loan of $1$ is $1/a(12H,i/12)$ so the annual debt service or mortgage
constant $R_{m}$ is $12$ times the monthly payment for a loan of $1$. We
furthermore assume that the asset value depreciates exactly as the mortgage is
paid off so the resale value at the end of the holding period is $V-$ $MV$
(and there is no remaining mortgage to pay off). Hence the value equation is:

\begin{center}
$V=a\left(  H,Y\right)  \left[  NOI-MVR_{m}\right]  +\left(  1+Y\right)
^{-H}\left[  V-MV\right]  +$ $MV$
\end{center}

By comparing this value equation with the previous one, we see that the only
difference is that $i$ is replaced by $R_{m}$ so the direct capitalization
rate will be:

\begin{center}
$R=MR_{m}+\left(  1-M\right)  Y$.
\end{center}

\subsection{Ellwood and Akerson Formulas with Constant Income}

We now consider a more general case where the mortgage is amortized over a
period longer than the holding period. With monthly payments, the mortgage
constant $R_{m}$ is $12$ times the monthly payment and the balance due on the
mortgage at the end of the holding period is $MVbal(12H)$. We further assume
that the asset appreciates by the proportion $\Delta_{0}$ over the holding
period so the resale value is $(1+\Delta_{0})V$. These assumptions yield the
value equation:

\begin{center}
$V=a\left(  H,Y\right)  \left[  NOI-MVR_{m}\right]  +\left(  1+Y\right)
^{-H}\left[  \left(  1+\Delta_{0}\right)  V-MVbal\left(  12H\right)  \right]
+$ $MV$
\end{center}

Collecting the $V$ terms to the left yields:

\begin{center}
$V\left[  1+a\left(  H,Y\right)  R_{m}M-\frac{1+\Delta_{0}}{\left(
1+Y\right)  ^{H}}+\frac{Mbal\left(  12H\right)  }{\left(  1+Y\right)  ^{H}%
}-M\right]  =a\left(  H,Y\right)  NOI$
\end{center}

Dividing through and rearranging terms gives:

\begin{center}
$V=\frac{NOI}{\left[  \frac{1}{a\left(  H,Y\right)  }+R_{m}M-\frac
{1+\Delta_{0}}{a\left(  H,Y\right)  \left(  1+Y\right)  ^{H}}+\frac
{Mbal\left(  12H\right)  }{a\left(  H,Y\right)  \left(  1+Y\right)  ^{H}%
}-\frac{M}{a\left(  H,Y\right)  }\right]  }$.
\end{center}

The denominator is the direct capitalization rate $R$. We can then use the
previous equations

\begin{center}
$s\left(  H,Y\right)  =a\left(  H,Y\right)  \left(  1+Y\right)  ^{H}$ and
$\frac{1}{a\left(  H,Y\right)  }=Y+\frac{1}{s\left(  H,Y\right)  }$
\end{center}

\noindent to simplify the rate $R$ to:

\begin{center}
$R=Y+\frac{1}{s\left(  H,Y\right)  }+R_{m}M-\frac{1}{s\left(  H,Y\right)
}-\frac{\Delta_{0}}{s\left(  H,Y\right)  }+\frac{Mbal\left(  12H\right)
}{s\left(  H,Y\right)  }-M\cdot Y-\frac{M}{s\left(  H,Y\right)  }$.
\end{center}

Canceling terms and using the equations $SFF(H,Y)=1/s(H,Y)$ and
$P=P(12H)=1-bal(12H)$ we can simply the expression to:

\begin{center}
$R=Y-M\left[  Y+PSFF\left(  H,Y\right)  -R_{m}\right]  -\Delta_{0}SFF\left(
H,Y\right)  $.
\end{center}

The expression in the square brackets is called the \textit{Ellwood
}$\mathit{C}$\textit{ factor} so the direct capitalization rate can be written
in the \textit{Ellwood form} as:

\begin{center}
$R=Y-MC-\Delta_{0}SFF\left(  H,Y\right)  $

Ellwood Formula
\end{center}

\noindent where $C=Y+PSFF(H,Y)-R_{m}$.

If we regroup the terms in another way reminiscent of the band of investment
formula than we have the:

\begin{center}
$R=MR_{m}+\left(  1-M\right)  Y-MPSFF\left(  H,Y\right)  -\Delta_{0}SFF\left(
H,Y\right)  $

Akerson Formula.
\end{center}

\section{The Valuation of Changing Income Streams}

\subsection{Introduction}

There is, of course, a general formula for the value $V$ of any income stream
$I_{1},I_{2},...,I_{n}$:

\begin{center}
$V=\sum_{k=1}^{n}\frac{I_{k}}{\left(  1+i\right)  ^{k}}$
\end{center}

\noindent but it is in fact the definition of the present value of the income
stream. We will consider changing income streams where the $I_{k}$'s are
defined in a regular manner by some relationship, and then we will seek a
concise formula for the above defined value $V$ (that is not just the defining
summation of the present values). These concise formulas are of more
theoretical than practical importance in the sense that an appraiser equipped
with an electronic spreadsheet can now directly use the definition to arrive
at a numerical value for the present value of a projected numerical income stream.

We will present a formula for the valuation of changing income streams defined
by linear recurrence relations (linear difference equations) which seems to be
new and to have all the usual formulas for valuing regular income streams as
special cases (e.g., straight line changing annuity, exponential or constant
ratio changing annuity, and streams changing according to the Ellwood $J$ premise).

As a special application, we show that the straight line and Hoskold methods
of capitalizing income streams can be seen as the discounted present value of
declining streams where the decline in income can be conceptualized as
interest losses. These losses result, as it were, from a make-believe
reinvestment of a capital recovery portion of the income in a hypothetical
sinking fund with an interest rate below the discount rate ($0$ in the
straight line case and some "safe" rate is in the Hoskold case). The declining
income stream of the straight line case can be evaluated using a known formula
for the straight line changing annuity. The more general formula given here is
needed for the declining income stream of the Hoskold case.

\subsection{Valuing Income Streams Defined by Linear Recurrence Relations}

\subsubsection{The general linear recurrence formulas}

Consider the general linear recurrence relation defined by

\begin{center}
$y_{0}=c$ and $y_{k}=my_{k-1}+b$
\end{center}

\noindent for some constants $m$, $b$, and $c$.

The general solution has the form

\begin{center}
$y_{n}=m^{n}c+m^{n-1}b+...+mb+b$
\end{center}

\noindent which can be expressed by the formula:

\begin{center}
$y_{n}=\left\{
\begin{array}
[c]{c}%
m^{n}c+\frac{b\left[  m^{n}-1\right]  }{m-1}\text{ for }m\neq1\\
c+nb\text{ for }m=1
\end{array}
\text{.}\right.  $
\end{center}

Taking the $k^{th}$ year's income as $y_{k}$ for $k=1,...,n$, the present
value of the income stream is;

\begin{center}
$V_{n}=\sum_{k=1}^{n}\frac{y_{k}}{\left(  1+i\right)  ^{k}}$.
\end{center}

It will be useful to notice the recurrence relation for the $V_{k}$'s:

\begin{center}
$V_{k}=\frac{m}{1+i}\left[  V_{k-1}+c\right]  +ba\left(  k,i\right)  $.
\end{center}

In Appendix $1$, we derive the formula for $V_{n}$ in the following four cases
where we use the notation $a_{n}=a(n,i)$. Since the $y_{k}$'s are defined by
general linear recurrence relations, we will call the formulas the
\textit{general linear recurrence valuation formulas}:

Case 1 for $m\neq1$, $\neq1+i$:$\qquad\qquad V_{n}=\left[  \frac{b}%
{m-1}+c\right]  \frac{m\left[  1-\left(  \frac{m}{1+i}\right)  ^{n}\right]
}{1+i-m}-\frac{b}{m-1}a_{n}$

Case 2 for $m=1$, $i\neq1$: $\qquad\qquad V_{n}=nc+\frac{b\left[
n-a_{n}\right]  }{i}$

Case 3 for $m=1,i\neq0$: \qquad$\qquad V_{n}=\left[  c+\left(  n+1\right)
b\right]  a_{n}-\frac{b\left[  n-a_{n}\right]  }{i}$

Case 4 for $m=1,i=0$: \qquad$\qquad V_{n}=nc+\frac{bn\left(  n+1\right)  }{2}$.

%\qquad

Real estate appraisal often considers an income stream of the special form

\begin{center}
$d,d-y_{1}h,d-y_{2}h,...,d-y_{n-1}h$
\end{center}

\noindent for constants $d$ and $h$. The stream has the present value:

\begin{center}
$V^{\ast}=\frac{d}{\left(  1+i\right)  ^{1}}+\frac{d-y_{1}h}{\left(
1+i\right)  ^{2}}+\cdots+\frac{d-y_{n-1}h}{\left(  1+i\right)  ^{n}}%
=da_{n}-\frac{h}{1+i}V_{n-1}$.
\end{center}

Using the recurrence relation for the $V_{k}$'s, we have:

\begin{center}
$V^{\ast}=da_{n}-\frac{h}{1+i}\left[  V_{n}-ba_{n}-\frac{mc}{1+i}\right]
\frac{1+i}{m}$
\end{center}

\noindent which simplifies to the formula for $V^{\ast}$ in terms of $V_{n}$
which, in turn, can be evaluated in the previous four cases:

\begin{center}
$V^{\ast}=\left[  d+\frac{bh}{m}\right]  a_{n}-\frac{hV_{n}}{m}+\frac{hc}%
{1+i}$.
\end{center}

\subsubsection{Application 1: The Straight Line Changing Annuity Formula}

The formula for valuing the linear changing annuity stream
$d,d-h,d-2h,...,d-(n-1)h$ can be obtained by taking $m=b=1$ and $c=0$ so that
$y_{k}=k$. Using the previous formula of $V^{\ast}$ and $V_{n}$ in case $3$
when $m=1\neq1+i$, we have:

\begin{center}
$V^{\ast}=\sum_{k=1}^{n}\frac{d-\left(  k-1\right)  h}{\left(  1+i\right)
^{k}}=\left[  d+h\right]  a_{n}-hV_{n}$

$=\left[  d+h\right]  a_{n}-h\left(  n+1\right)  a_{n}+\frac{h\left[
n-a_{n}\right]  }{i}$

$=\left[  d-nh\right]  a_{n}+\frac{h\left[  n-a_{n}\right]  }{i}$
\end{center}

\noindent which was the previously known formula for valuing the straight line
(constant amount) changing income stream.

\subsubsection{Application 2: The Constant Ratio Changing Annuity Formula}

Suppose an income stream starts with $1$ at the end of year one and then grows
at a rate of $g$ for $n$ years. To apply the general formula, take $b=0$ and
$m=1+g$. In order to start with $y_{1}=1$, we must take $y_{0}=c=1/(1+g)$ so
that $y_{k}=(1+g)^{k-1}$. Using the general formula in case $1$, we have:

\begin{center}
$V_{n}=\frac{1}{1+g}\frac{\left(  1+g\right)  \left[  1-\left(  \frac
{1+g}{1+i}\right)  ^{n}\right]  }{i-g}=\frac{\left[  1-\left(  \frac{1+g}%
{1+i}\right)  ^{n}\right]  }{i-g}$
\end{center}

\noindent which is the usual formula for evaluating the constant ratio
changing annuity.

\subsubsection{Application 3: The Ellwood J Factor and Ellwood R Formulas}

Recall that

\begin{center}
$s_{n}=s\left(  n,i\right)  =\left(  1+i\right)  ^{n-1}+\left(  1+i\right)
^{n-2}+\cdots+\left(  1+i\right)  ^{1}+1$

$=\frac{\left(  1+i\right)  ^{n}-1}{i}=\left(  1+i\right)  ^{n}a_{n}=\frac
{1}{SFF\left(  n,i\right)  }$
\end{center}

\noindent is the accumulation of one per period. It is useful to first use the
general formula to derive the value of the stream of incomes $s_{1}%
,s_{2},...,s_{n}$ at the end of years $1,2,...,n$. In this case, $m=1+i,b=1$,
and $c=0$. Then the formula yields in case $2$:

\begin{center}
$\sum_{k=1}^{n}\frac{s_{k}}{\left(  1+i\right)  ^{k}}=\sum_{k=1}^{n}%
a_{k}=\frac{n-a_{n}}{i}$.
\end{center}

The Ellwood $J$ premise is that the income stream will change by an amount
$\Delta I$ over $n$ years after starting with a (hypothetical) value at time
$0$ of $I$ (where $\Delta$ is the relative change in $I$). The change,
however, occurs in a particular way. At the end of the $k^{th}$ year, the
income is $I+s_{k}h$ for some fixed $h$. Since we must have the income at the
end of the nth year as $I+s_{n}h=I+\Delta I$ we can quickly solve for $h$ as
$h=\Delta I/s_{n}$. The actual income stream starts at the end of year $1$ so
its value is:

\begin{center}
$V^{\ast}=\sum_{k=1}^{n}\frac{I+s_{k}h}{\left(  1+i\right)  ^{k}}=Ia_{n}%
+h\sum_{k=1}^{n}\frac{s_{k}}{\left(  1+i\right)  ^{k}}$.
\end{center}

Using the previous formula for the present value of $s_{k}$'s income steam and
the definition of $h$, we have:

\begin{center}
$V^{\ast}=Ia_{n}+\frac{\Delta I\left[  n-a_{n}\right]  }{s_{n}i}=I\left[
a_{n}+\frac{\Delta}{s_{n}}\frac{\left[  n-a_{n}\right]  }{i}\right]  $
\end{center}

\noindent so the reciprocal of the term in the square brackets is the
capitalization rate $R$ that would yield the value as $V^{\ast}=I/R$. The cap
rate $R$ can then be simplified as follows.

\begin{center}
$R=\frac{1}{\left[  a_{n}+\frac{\Delta}{s_{n}}\frac{\left[  n-a_{n}\right]
}{i}\right]  }=\frac{1-\left(  1+i\right)  ^{-n}+\left(  1+i\right)  ^{-n}%
}{\left[  a_{n}+\frac{\Delta na_{_{n}}}{s_{n}\left[  1-\left(  1+i\right)
^{-n}\right]  }-\frac{\Delta a_{n}}{s_{n}i}\right]  }=\frac{ia_{n}+a_{n}%
/s_{n}}{a_{n}\left\{  1+\Delta\left[  \frac{1}{s_{n}}\left(  \frac
{n}{1-\left(  1+i\right)  ^{-n}}-\frac{1}{i}\right)  \right]  \right\}  }$
\end{center}

Thus the capitalization rate $R$ can be simplified to:

\begin{center}
$R=\frac{i+1/s_{n}}{1+\Delta J}$
\end{center}

\noindent where

\begin{center}
$J=\frac{1}{s_{n}}\left(  \frac{n}{1-\left(  1+i\right)  ^{-n}}-\frac{1}%
{i}\right)  $
\end{center}

\noindent is the Ellwood $J$ factor. We have only been considering income
streams defined by certain formulas. Thus we have not considered any extra
term at the end of year $n$ for the terminal value of some underlying asset.
In other words we are assuming that any underlying asset wastes away to value
zero at the end of year $n$. Otherwise, the "$1$" in the numerator of the
expression for $R$ would be replaced by the relative drop $\Delta_{0}$ in the
overall value of the asset ($\Delta_{0}=1$ in our case).

Our previous presentation of the Ellwood mortgage analysis with a constant
income stream can now be easily modified to accommodate an income stream
changing according to the Ellwood $J$ premise used above. Carrying over the
relevant notation from our previous mortgage analysis, the value equation is:

\begin{center}
$V=\sum_{k=1}^{H}\frac{I+s_{k}h-R_{m}MV}{\left(  1+Y\right)  ^{k}}+\left(
1+Y\right)  ^{-H}\left[  \left(  1+\Delta_{0}\right)  V-MVbal\left(
12H\right)  \right]  +$ $MV$
\end{center}

\noindent where $s_{k}=s(k,Y)$ and $h=I\Delta/sH$. Using the previous result:

\begin{center}
$\sum_{k=1}^{H}\frac{s_{k}}{\left(  1+Y\right)  ^{k}}=\frac{H-a_{H}}{Y}$
\end{center}

\noindent where $a_{H}=a(H,Y)$, the value equation can be simplified to:

\begin{center}
$V=Ia_{H}-R_{m}$ $MVa_{H}+\frac{I\Delta\left[  H-a_{H}\right]  }{s_{H}%
Y}+\left(  1+Y\right)  ^{-H}\left[  \left(  1+\Delta_{0}\right)
V-MVbal\left(  12H\right)  \right]  +$ $MV$.
\end{center}

Collecting the $V$ terms on the left-hand side yields:

\begin{center}
$V\left[  1+R_{m}MVa_{H}-\frac{1+\Delta_{0}}{\left(  1+Y\right)  ^{-H}}%
-\frac{Mbal\left(  12H\right)  }{\left(  1+Y\right)  ^{-H}}-M\right]
=I\left[  a_{H}+\frac{\Delta\left[  H-a_{H}\right]  }{s_{H}Y}\right]  $
\end{center}

Then we can skip some algebra since the square brackets on the left-hand side
are developed exactly as in the previous treatment of the Ellwood mortgage
analysis and the square brackets on the right-hand side are developed like the
treatment of Ellwood $J$ factor above. Thus we can quickly arrive at the
$V=I/R$ formula with:

\begin{center}
$R=\frac{Y-MC-\Delta_{0}SFF}{1+\Delta J}$

Ellwood's $R$ with Changes in Income and Asset Value
\end{center}

\noindent where Ellwood's $C=Y+PSFF-R_{m}$ as before and $SFF=SFF(H,Y)=1/s_{H}%
$.

\subsubsection{The Straight Line and Hoskold Capitalization Rates}

There is some controversy in the field of real estate appraisal over the
status of the so-called "straight line" method (also called "Ring" method) and
the Hoskold method of determining direct capitalization rates. See Table 3.

\begin{center}%
\begin{tabular}
[c]{cccc}%
Method to determine &  &  & \\
capitalization rate & Return on investment & +Return on investment & =
Capitalization rate $R$\\\hline
\multicolumn{1}{|c}{Straight line method} & \multicolumn{1}{|c}{$SFF\left(
n,i\right)  $} & \multicolumn{1}{|c}{$i$} & \multicolumn{1}{|c|}{$i+1/n$%
}\\\hline
\multicolumn{1}{|c}{Hoskold method @ $i_{S}$} &
\multicolumn{1}{|c}{$SFF\left(  n,i_{S}\right)  $} & \multicolumn{1}{|c}{$i$}
& \multicolumn{1}{|c|}{$i+SFF\left(  n,i_{S}\right)  $}\\\hline
\multicolumn{1}{|c}{Annuity method @ $i$} & \multicolumn{1}{|c}{$SFF\left(
n,i\right)  $} & \multicolumn{1}{|c}{$i$} & \multicolumn{1}{|c|}{$1/a\left(
n,i\right)  $}\\\hline
\end{tabular}

Table 3: Methods to give cap rates
\end{center}

We will show that the straight line and Hoskold capitalization rates will,
when divided into the first year's income, give the correct present value for
certain declining income streams.

\subsubsection{The Straight Line Capitalization Formula}

We will show that the straight line formula (as well as the Hoskold formula)
applies to certain declining income streams from an income property (without
any reference to a sinking fund). Sinking funds are relevant as a heuristic
device because one can "motivate" the declining income stream as the combined
income stream yielded by the composite investment of an income property giving
a level income stream plus a sinking fund with a sub-standard interest rate.
The decline in the total or composite income stream is precisely equal to the
interest rate losses due to the reinvestment at a substandard interest rate.
This sinking fund would usually be a hypothetical or "as if" device. The
decline in the income stream is "as if" part of the proceeds of a level stream
were reinvested at a "safe" rate below the prevailing interest rate.

Consider a declining income stream with $d$ as the first year's income which
then declines by the amount $h$ each year for $n$ years. The present value of
the income stream at the discount rate $i$ is:

\begin{center}
$V=\frac{d}{\left(  1+i\right)  ^{1}}+\frac{d-h}{\left(  1+i\right)  ^{2}%
}+\frac{d-2h}{\left(  1+i\right)  ^{3}}+\cdots+\frac{d-\left(  n-1\right)
h}{\left(  1+i\right)  ^{n}}$.
\end{center}

The straight line changing annuity formula for this sum was previously derived.

\begin{center}
$V=\left[  d-nh\right]  a\left(  n,i\right)  +\frac{h\left[  n-a\left(
n,i\right)  \right]  }{i}$.
\end{center}

\noindent The formula can, of course, be applied as well to straight line
rising income streams by considering $h$ as being negative.

The straight line capitalization formula can be obtained as a special case. We
consider the hypothetical composite investment consisting of an income
property with level income $I$ and reinvest of the capital recovery portion of
income in a mattress sinking fund. Suppose that the income only from a
property is constant amount $I$ for $n$ years. At the end of each year part of
the proceeds are reinvested in a sinking fund at the ultra-safe or "mattress"
interest rate of zero. The value of the composite investment, property plus
sinking fund, is $V$. At the end of each year, $SFF(n,0)V=V/n$ is invested in
the zero-interest sinking fund. Thus at the end of second year, there is an
interest loss of $h=iV/n$. At the end of each subsequent year, there is an
additional loss of $h=iV/n$. Thus the combined income stream is precisely of
the straight line changing annuity kind with $d=I$ and $h=iV/n$. Applying the
valuation formula, we have:

\begin{center}
$V=\left(  I-n\frac{iV}{n}\right)  a\left(  n,i\right)  +\frac{\frac{iV}%
{n}\left(  n-a\left(  n,i\right)  \right)  }{i}=Ia\left(  n,i\right)
-iVa\left(  n,i\right)  +V-Va\left(  n,i\right)  /n$
\end{center}

Solving for $V$ yields the straight line formula:

\begin{center}
$V=\frac{Ia\left(  n,i\right)  }{\left(  i+1/n\right)  a\left(  n,i\right)
}=\frac{I}{i+1/n}=\frac{I}{i+SFF\left(  n,0\right)  }$.

Straight Line Capitalization Formula
\end{center}

Thus the specific declining income stream appropriate for the straight line
formula can be motivated as the composite result of a constant income stream
plus reinvestment of part of the proceeds each year in a mattress sinking
fund. It is unlikely that an appraiser will be asked to appraise the composite
investment of a level income property plus a mattress sinking fund. Thus it is
easy to see that the sinking fund in this case is only a heuristic or
hypothetical device to motivate the decline in the income stream "as if" they
were the interest losses from a mattress sinking fund. The sinking fund is
just as hypothetical in the Hoskold case.

\subsubsection{The Hoskold Formula}

We must use case $1$ in our more general valuation formula to evaluate the
declining income stream that underlies the Hoskold formula. We will show that
the Hoskold formula works for a certain declining income stream

\begin{center}
$I,I-y_{1}h,I-y_{2}h,...,I-y_{n-1}h$
\end{center}

\noindent where $m=1+i_{s}$, $b=1$, and $c=0$, and where is is a "safe"
interest rate intermediate between $i$ and $0$. Then using the previous
formula for $V^{\ast}$ with $d=I$, we have:

\begin{center}
$V^{\ast}=\left[  I+\frac{h}{1+i_{S}}\right]  a_{n}-\frac{hV_{n}}{1+i_{S}}$
\end{center}

\noindent so substituting in the formula for $V_{n}$ (case 1 of $m\neq1,1+i$)
yields after some algebra:

\begin{center}
$V^{\ast}=Ia_{n}-\frac{h}{i_{S}}\left[  \frac{1-\left(  \frac{1+i_{S}}%
{1+i}\right)  ^{n}}{i-i_{S}}-a_{n}\right]  $

$V^{\ast}$ Formula in Hoskold Case
\end{center}

To arrive at the specific declining income stream for the Hoskold case, we
must fix $h$ as the interest loss resulting from investing in the sub-standard
sinking fund at the safe rate $i_{s}$. The declining stream is then motivated
as the composite result of a constant income stream at the level $d$ minus the
interest losses in the safe sinking fund. The term subtracted from $d$ in year
$k+1$ for $k=1,...,k-1$ is $y_{k}h$. Remembering that $m=1+i_{s},b=1$, and
$c=0$ in this Hoskold case, the $y_{k}$ term is:

\begin{center}
$y_{k}=\left(  1+i_{S}\right)  ^{k-1}+\cdots+\left(  1+i_{S}\right)
+1=s\left(  k,i_{S}\right)  =\frac{\left(  1+i_{S}\right)  ^{k}-1}{i_{S}%
}=\frac{1}{SFF\left(  k,i_{S}\right)  }$
\end{center}

\noindent where the sinking fund factor $SFF(k,i_{s})$ is the amount invested
at the end of each year for $k$ years to accumulate to $1$ at the end of year
$k$ at the interest rate $i_{s}$. In our safe sinking fund, we must invest at
the end of each year for $n$ years the amount that will accumulate to
$V^{\ast}$, and that amount is $V^{\ast}SFF(n,i_{s})$. After that amount
$i_{s}$ invested at the end of year $1$, the interest rate loss at the end of
year $2$ from investing in the substandard sinking fund is $(i-i_{s})V^{\ast
}SFF(n,i_{s})$ which should equal $y_{1}h$. At the end of year $3$, there is
the same loss on the amount invested at the end of year $2$ but there is also
the loss of what would have been the sinking fund accumulation on the previous
loss. Thus the loss at the end of year $3$ is:

\begin{center}
$\left[  \left(  1+i_{S}\right)  +1\right]  \left(  i-i_{S}\right)  V^{\ast
}SFF\left(  n,i_{S}\right)  =\left(  i-i_{S}\right)  V^{\ast}SFF\left(
n,i_{S}\right)  s\left(  2,i_{S}\right)  =y_{2}h$.
\end{center}

By similar reasoning we see that the loss at the end of year $k+1$ is:

\begin{center}
$\left(  i-i_{S}\right)  V^{\ast}SFF\left(  n,i_{S}\right)  s\left(
k,i_{S}\right)  =y_{k}h$.
\end{center}

Since we know that $y_{k}=s(k,i_{s})$, we see that:

\begin{center}
$h=\left(  i+i_{S}\right)  V^{\ast}SFF\left(  n,i_{S}\right)  =\frac{\left(
i-i_{S}\right)  V^{\ast}i_{S}}{\left(  1+i_{S}\right)  ^{n}-1}$
\end{center}

\noindent in the formula for $V^{\ast}$ in the Hoskold case.

Substituting $h$ into the $V^{\ast}$ formula for the Hoskold case yields:

\begin{center}
$V^{\ast}=Ia_{n}-\frac{h}{i_{S}}\left[  \frac{1-\left(  \frac{1+i_{S}}%
{1+i}\right)  ^{n}}{i-i_{S}}-a_{n}\right]  =Ia_{n}-\frac{\left(
i-i_{S}\right)  V^{\ast}SFF\left(  n,i_{S}\right)  }{i_{S}}\left[
\frac{1-\left(  \frac{1+i_{S}}{1+i}\right)  ^{n}}{i-i_{S}}-a_{n}\right]  $
\end{center}

\noindent which simplifies to:

\begin{center}
$V^{\ast}=Ia_{n}+\frac{\left[  \left(  \frac{1+i_{S}}{1+i}\right)
^{n}-\left(  \frac{1}{1+i}\right)  ^{n}\right]  V^{\ast}}{\left(
1+i_{S}\right)  ^{n}-1}-a_{n}V^{\ast}SFF\left(  n,i_{S}\right)  $.
\end{center}

Collecting all the $V^{\ast}$ terms on the left side yields:

\begin{center}
$V^{\ast}\left[  \frac{(1+i_{S})^{n}-1-\left(  \frac{1+i_{S}}{1+i}\right)
^{n}+\left(  \frac{1}{1+i}\right)  ^{n}}{\left(  1+i_{S}\right)  ^{n}%
-1}\right]  +a_{n}V^{\ast}SFF\left(  n,i_{S}\right)  =Ia_{n}$
\end{center}

\noindent where the term in the square brackets simplifies to:

\begin{center}
$\frac{\left(  \left(  1+i_{S}\right)  ^{n}-1\right)  \left(  1-\left(
\frac{1}{1+i}\right)  ^{n}\right)  }{\left(  1+i_{S}\right)  ^{n}-1}=1-\left(
\frac{1}{1+i}\right)  ^{n}=ia_{n}$.
\end{center}

Therefore we have $V^{\ast}[i+SFF(n,i_{s})]a_{n}=Ia_{n}$ so we can cancel
$a_{n}$ and solve for the value $V^{\ast}$ of the declining income stream
$I,I-y_{1}h,...,I-y_{n}-1h$ (with $m=1+i_{s},b=1$, and $c=0$ in the definition
of $y_{k}$) as:

\begin{center}
$V^{\ast}=\frac{I}{i+SFF\left(  n,i_{S}\right)  }$.

The Hoskold Formula
\end{center}

\subsection{Generalized Amortization Tables: The Main Theorem}

We have relied mostly on the language of algebra. Since not all appraisers are
fluent in that language, it might be useful to restate some of the results
using amortization tables. We begin with a general result about amortization
tables where the principal reductions $P_{1}$, $P_{2}$, ..., $P_{n}$ are
arbitrarily given along with the interest or discount rate $i$. The value $V$
is the sum of the principal reductions. The incomes (or payments) per period
are determined from this data. The Main Theorem is that the discounted present
value of the incomes determined in this manner from the given $P_{k}$'s is the
value $V$ which is the sum of the $P_{k}$'s. For the results about the Ring
and Hoskold methods, we consider amortization tables where the principal
reductions or capital recovery entries are generated by a sinking fund at a
rate $r$ not necessarily the same as the discount rate $i$. When $r=0$, we
will have an amortization table for the straight line or Ring method which
shows the declining income for that case. When $r=i_{s}$ between $0$ and $i$,
we have a Hoskold amortization table that shows the declining income for that
case. When $r=i$, we have usual amortization table with level income or
amortization payments. If $r>i$, we have an amortization table with involves
capital recovery at a supra-standard rate $r$ and which thus generates a
rising income stream.

The principal or capital to be recovered is defined as the sum of those given
principal reductions. Certain relationships hold between the columns in an
amortization table. The interest in each year is the rate $i$ times the
balance or unrecovered capital from the previous year. The entry in the
payment or income column is the sum of the interest and principal reduction
(or capital recovery) columns. The entry in the balance (or unrecovered
capital) column is the previous entry in the column minus the principal
reduction (or capital recovery). The last entry in the balance or unrecovered
capital column is zero.

Let $P_{1},P_{2},...,P_{n}$ be the given principal reductions, let
$V=P_{1}+P_{2}+...+P_{n}$ be the sum, and let $i$ be the discount rate. That
is the only data given for the following general theorem about amortization tables.

\begin{center}%
\begin{tabular}
[c]{ccccc}\hline
\multicolumn{1}{|c}{Year} & \multicolumn{1}{|c}{Income} &
\multicolumn{1}{|c}{= Interest +} & \multicolumn{1}{|c}{Principal reduction} &
\multicolumn{1}{|c|}{Balance}\\\hline\hline
\multicolumn{1}{|c}{$1$} & \multicolumn{1}{|c}{$I_{1}=P_{1}+i(P_{1}%
+\cdots+P_{n})$} & \multicolumn{1}{|c}{$iV$} & \multicolumn{1}{|c}{$P_{1}$} &
\multicolumn{1}{|c|}{$V-P_{1}$}\\\hline
\multicolumn{1}{|c}{$2$} & \multicolumn{1}{|c}{$I_{2}=P_{2}+i(P_{2}%
+\cdots+P_{n})$} & \multicolumn{1}{|c}{$i\left(  V-P_{1}\right)  $} &
\multicolumn{1}{|c}{$P_{2}$} & \multicolumn{1}{|c|}{$V-P_{1}-P_{2}$}\\\hline
\multicolumn{1}{|c}{$\cdots$} & \multicolumn{1}{|c}{$\cdots$} &
\multicolumn{1}{|c}{$\cdots$} & \multicolumn{1}{|c}{$\cdots$} &
\multicolumn{1}{|c|}{$\cdots$}\\\hline
\multicolumn{1}{|c}{$k$} & \multicolumn{1}{|c}{$I_{k}=P_{k}+i(P_{k}%
+\cdots+P_{n})$} & \multicolumn{1}{|c}{$i\left(  V-P_{1}-...-P_{k-1}\right)
$} & \multicolumn{1}{|c}{$P_{k}$} & \multicolumn{1}{|c|}{$V-P_{1}-...-P_{k}$%
}\\\hline
\multicolumn{1}{|c}{$\cdots$} & \multicolumn{1}{|c}{$\cdots$} &
\multicolumn{1}{|c}{$\cdots$} & \multicolumn{1}{|c}{$\cdots$} &
\multicolumn{1}{|c|}{$\cdots$}\\\hline
\multicolumn{1}{|c}{$n$} & \multicolumn{1}{|c}{$I_{n}=P_{n}+iP_{n}$} &
\multicolumn{1}{|c}{$i\left(  V-P_{1}-...-P_{n-1}\right)  $} &
\multicolumn{1}{|c}{$P_{n}$} & \multicolumn{1}{|c|}{$V-\sum_{k=1}^{n}P_{k}=0$%
}\\\hline
&  &  & $\sum_{k=1}^{n}P_{k}=V$ &
\end{tabular}

Table 4: General Amortization Table
\end{center}

The other columns are all defined in terms of the given $P_{i}$'s in the
manner indicated. The incomes $I_{k}$'s are determined as the sum of the
Interest and Principal Reduction columns, and the general formula is

\begin{center}
$I_{k}=P_{k}+i(P_{k}+...+P_{n})$.
\end{center}

The Main Theorem is that the discounted present value of these incomes is the
value $V$, the sum of the arbitrarily given $P_{k}$'s.

\begin{center}
$\sum_{k=1}^{n}\frac{I_{k}}{\left(  1+i\right)  ^{k}}=\sum_{k=1}^{n}P_{k}$

Main Theorem on Amortization Tables
\end{center}

\noindent The proof in given in Appendix $2$.

\subsection{Amortization Tables with Sinking Fund Capital Recovery}

Let $V$ be the value of the investment (or loan) and $n$ the number of years
to recover the capital (or pay off the loan). Let $i$ be the interest rate and
$r$ be the rate for the capital recovery sinking fund. The value of the first
year's income (or payment) is $I$. The value $V$ is related to the first
year's income by the direct capitalization formula:

\begin{center}
$V=\frac{I}{i+SFF\left(  n.r\right)  }$..
\end{center}

The new deposit in the sinking fund each year to recover the capital is
$SFF(n,r)V$ which is abbreviated $SFFV$. After the deposit at the end of the
$k^{th}$ year, the amount in the sinking fund is $SFFVs(k,r)$ which
abbreviated $SFFVs_{k}$. Therefore the capital recovery during the $k^{th}$
year due to both the new deposit and the new interest is $SFFVs_{k}%
-SFFVs_{k-1}=SFFV(1+r)^{k-1}$ and that is the entry in the $k^{th}$ row of the
capital recovery (or principal reduction) column. Each year's income $I_{k}$
beginning with $I_{1}=I$ is the sum of the interest (or return on unrecovered
capital) and the capital recovered (return of capital) for that year.

\begin{center}%
\begin{tabular}
[c]{|c|c|c|c|c|}\hline
Year & Income & = Interest + & Capital Recovered & Balance\\\hline\hline
$1$ & $I$ & $iV$ & $SFFV$ & $V\left(  1-SFF\right)  $\\\hline
$2$ & $I_{2}$ & $iV\left(  1-SFF\right)  $ & $SFFV\left(  1+r\right)  $ &
$V\left(  1-SFFs_{2}\right)  $\\\hline
$3$ & $I_{3}$ & $iV\left(  1-SFFs_{2}\right)  $ & $SFFV\left(  1+r\right)
^{2}$ & $V\left(  1-SFFs_{3}\right)  $\\\hline
$\cdots$ & $\cdots$ & $\cdots$ & $\cdots$ & $\cdots$\\\hline
$n$ & $I_{n}$ & $iV\left(  1-SFFs_{n-1}\right)  $ & $SFFV\left(  1+r\right)
^{n-1}$ & $V\left(  1-SFFs_{n}\right)  $\\\hline
\end{tabular}

Table 5: Amortization Table with Sinking Fund Capital Recovery
\end{center}

Since $SFF=1/s_{n}$ the last entry in the Balance or Unrecovered Capital
column is $0$. The sum of the Capital Recovered column is:

\begin{center}
$SFFV+SFFV\left(  1+r\right)  +SFFV\left(  1+r\right)  ^{2}+...+SFFV\left(
1+r\right)  ^{n-1}=SFFVs_{n}=V$
\end{center}

\noindent as desired. The incomes $I_{k}$ are obtained as the sum of the
Interest and Capital Recovered columns. It is useful to compute the first few incomes.

\begin{center}
$I_{2}=iV\left(  1-SFF\right)  +SFFV\left(  1+r\right)
=iV+SFFV-iSFFV+rSFFV=I-\left(  i-r\right)  SFFV$.
\end{center}

The income for the $2^{nd}$ year is $I$ minus $(i-r)SFFV$ which is the
interest loss on the sinking fund deposit of $SFFV$.

The third year's income is calculated as follows.

\begin{center}
$I_{3}=iV\left(  1-SFFs_{2}\right)  +SFFV\left(  1+r\right)  ^{2}%
=iV-iVSFF+SFFV\left(  1+r\right)  -\left(  i-r\right)  SFFV\left(  1+r\right)
$

$=I_{2}-\left(  i-r\right)  SFFV\left(  1+r\right)  =I-\left(  i-r\right)
SFFVs_{2}$.
\end{center}

Thus we see that each year's income $I_{k}$ is $I$ minus the interest losses
on the sinking fund (assuming $r<i$) where the latter can be calculated as
$(i-r)SFFVs_{k-1}$, the accumulation $s_{k}$ on the interest losses $(i-r)$ on
the sinking fund deposits $SFFV$:

\begin{center}
$I_{k}=I-(i-r)SFFVs_{k-1}.$
\end{center}

Since these incomes $I_{k}$ are the same as those obtained in our previous
analysis of the Hoskold case, the Main Theorem on Amortization Tables now
gives us another proof that the present value of these incomes is the value
$V=I/[i+SFF(n,r)]$ when $r=i_{s}$.

\begin{figure}[h]
\centering
\includegraphics[width=0.7\linewidth]{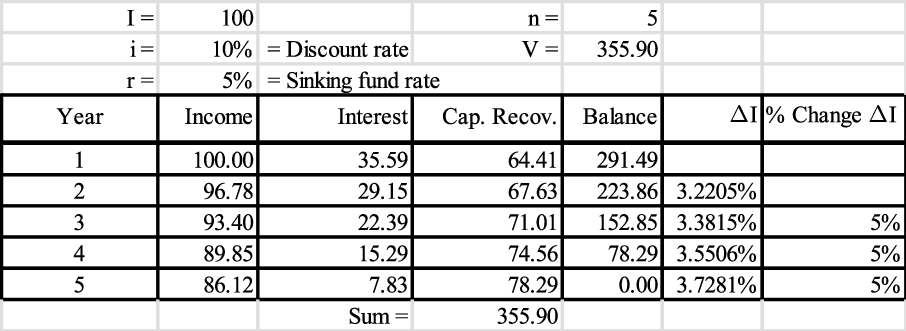}
\caption{Amortization table: Discount = 10\%, Sinking fund = 5\%}%
\label{fig:amort-table-10-5percent}%
\end{figure}

In the straight line or Ring case of , $SFF=1/n$ and $s_{k}=k$ so the
declining income is given by $I_{k}=I-i(V/n)k$. The income stream declines by
a constant amount $iV/n$ each year independent of $k$. In the Hoskold case,
the drop in the income stream from $I_{k}$ to $I_{k+1}$ is $(i-r)SFFV(s_{k}%
+1-s_{k})=(i-r)SFFV(1+r)^{k}$ which depends on $k$. Thus the Hoskold requires
the formula more general than the constant amount changing annuity formula.
The drop in the income stream in each period is $(1+r)$ times the previous
drop. This is illustrated in the following table based on the Hoskold
situation where $0<r<i$. The change in income accelerates at the sinking fund
rate of $r$ (as we see in the right-most column of the spreadsheet) as in
Figure \ref{fig:amort-table-10-5percent}.

\begin{figure}[h]
\centering
\includegraphics[width=0.7\linewidth]{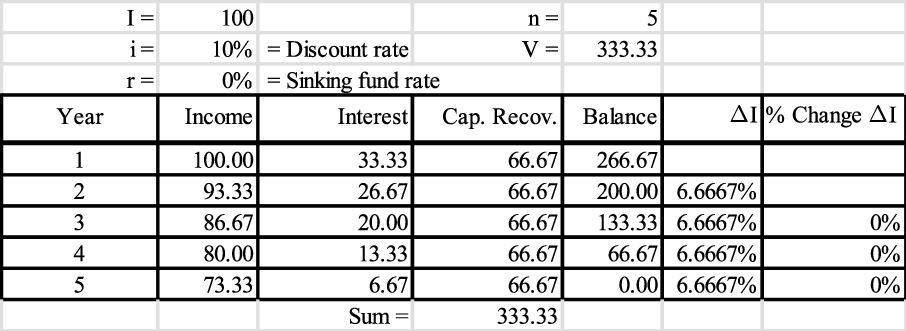}
\caption{Amortization table: Discount = 10\%, Sinking fund = 0\%}%
\label{fig:amort-table-10-0percent}%
\end{figure}

In the straight line or Ring case, we set the sinking fund rate to $0$ as in
Figure \ref{fig:amort-table-10-0percent}.

When $r=i$, we have an ordinary amortization table where $i-r=0$ so the
interest loss is $0$ and the income is constant.

\begin{figure}[h]
\centering
\includegraphics[width=0.7\linewidth]{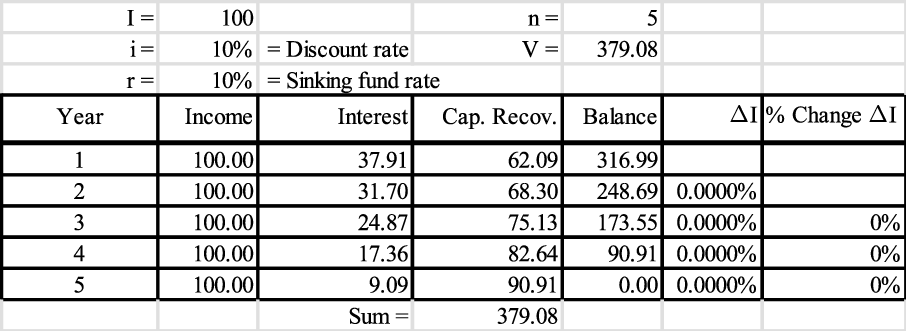}
\caption{Amortization table: Discount = 10\%, Sinking fund = 10\%}%
\label{fig:amort-table-10-10percent}%
\end{figure}

\section{The Internal Rate of Return}

\subsection{The Many Flaws and Few Benefits of IRR's}

What is the criteria to use to measure the benefits of an investment project?
It is the \textit{net present value} or $NPV$ of the project computed using a
discount rate appropriate for the riskiness of the project. There is an old
real estate saying that there are three things which determine the value of
real estate for retail purposes: location, location, and location. In a
similar manner, we can say there are three investment measuring devices:
$NPV$, $NPV$, and $NPV$. The internal rate of return or $IRR$ is not one of them.

Why analyze $IRR$ at all? The analysis of $IRR$ is important because it is
widely used by executives and consultants; \textquotedblleft as recently as
1999, academic research found that three-quarters of CFOs always or almost
always use IRR when evaluating capital projects.\textquotedblright%
\ \cite{kelleher:irr} However, many of those who recommend the $IRR$ concept
seem to be unaware or only vaguely aware of the many problems with $IRR$'s.
Hence it is necessary to reiterate the many fallacies in the use of $IRR$'s
and to show the limited domain where $IRR$'s can be correctly applied.

\subsection{Definition of IRR}

An \textit{investment project} is defined by a series of cash flows
$C_{0},C_{1},C_{2},...,C_{n},...$ where $C_{t}$ is the cashflow at the end of
time $t$ (time periods are taken as years). A negative cashflow $C_{t}$ is an
investment into the project and a positive cashflow $C_{t}$ is a payout from
the project. Given the discount rate $r$ (the opportunity cost of capital to
be invested in projects of similar riskiness), the net present value $NPV$ of
a project $C_{0},C_{1},C_{2},...,C_{n}$ is:

\begin{center}
$NPV(r)=\sum_{t=0}^{n}\frac{C_{t}}{\left(  1+r\right)  ^{t}}$
\end{center}

\noindent where we might write $NPV(r)$ to make explicit the use of $r$ as the
discount rate in the definition of $NPV$. An \textit{internal rate of return}
$IRR$ of the project can be defined as a rate which sets the net present value
to zero:

\begin{center}
$NPV(IRR)=\sum_{t=0}^{n}\frac{C_{t}}{\left(  1+IRR\right)  ^{t}}=0$.
\end{center}

\noindent While some may speak of \textquotedblleft the\textquotedblright%
\ $IRR$ of a project, there are some projects which have multiple $IRR$'s.

\begin{figure}[h]
	\centering
	\includegraphics[width=0.7\linewidth]{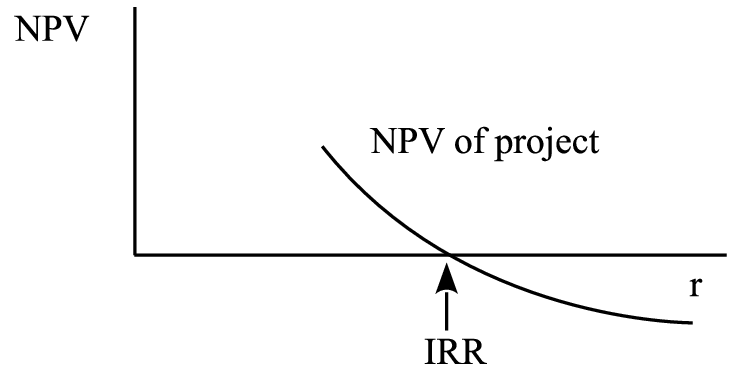}
	\caption{$IRR$ is the interest rate $r$ that sets $NPV\left(  r\right)  $ to $0$.}
	\label{fig:irr-pix}
\end{figure}

If we graph $NPV$ on the vertical axis and the discount rate $r$ on the
horizontal axis, then the $IRR$ is the discount rate at which the $NPV$ curve
cuts the horizontal axis as shown in Figure \ref{fig:irr-pix}.

\subsection{Examples of IRR's}

There is no simple formula for finding an $IRR$. Except in a few simple cases,
$IRR$'s (as the roots of a polynomial) are best computed through an iterative
procedure of ever closer approximation. Fortunately, such numerical
computational procedures are now built into most hand-held financial
calculators and spreadsheets so finding $IRR$'s is no longer a practical problem.

To construct an example with an $IRR=.20$ or $20\%$, choose any initial
investment of say $\$1000$ (so that $C_{0}=-1000$), and then take the
cashflows as the interest $\$200$ until the final time period when the
principal is returned as well. Table 8 gives the relevant information for
three projects.

\begin{center}%
\begin{tabular}
[c]{|c|c|c|c|c|c|c|c|}\hline
Project & $C_{0}$ & $C_{1}$ & $C_{2}$ & $C_{3}$ & $IRR$ & $NPV$ @ $10\%$ &
$NPV$ @ $12\%$\\\hline\hline
A & $-1000$ & $200$ & $200$ & $1200$ & $20\%$ & $\$248.69$ & $\$192.15$%
\\\hline
B & $-1000$ & $500$ & $500$ & $500$ & $23.38\%$ & $\$243.43$ & $\$200.92$%
\\\hline
C & $-1000$ & $120$ & $120$ & $120$ & $12\%$ & $\$49.74$ & $\$0$\\\hline
\end{tabular}

Table 8: Examples of Projects and their $IRR$'s.
\end{center}

\subsection{Pitfall 1 in Using IRR's: The Negative of a Project has the same
IRR}

One of the simplest \textquotedblleft rules\textquotedblright\ you will find
in the literature is that an investment project is profitable (that is, has
positive $NPV$) if its $IRR$ is greater than the interest rate $r$. But this
cannot be true without additional assumptions since the negative of a project
has the same $IRR$. Reversing all the cashflows reverses the role of the
lender and borrower. For instance consider the negative of project A as shown
in Table 9.

\begin{center}%
\begin{tabular}
[c]{|c|c|c|c|c|c|c|c|}\hline
Project & $C_{0}$ & $C_{1}$ & $C_{2}$ & $C_{3}$ & $IRR$ & $NPV$ @ $10\%$ &
$NPV$ @ $12\%$\\\hline\hline
-A & $1000$ & $-200$ & $-200$ & $-1200$ & $20\%$ & $-\$248.69$ &
$-\$192.15$\\\hline
\end{tabular}

Table 9: Project -A has same $IRR$ as Project A but with negative $NPV$ 's.
\end{center}

If the discount rate were, say, $10\%$ or $12\%$ then the project -A has a
greater$IRR$ of $20\%$ but a negative$NPV$ at those discount rates. In order
for $r<IRR$ to imply $0<NPV$, it is sufficient to assume that $NPV$ declines
as the discount rate increases, i.e., that the $NPV$ curve slopes downward
from left to right. Thus we have the rule:

\begin{center}
If the $NPV$ of a project declines as the discount rate $r$ increases then

$r<IRR$ implies $0<NPV$.
\end{center}

\subsection{Pitfall 2 in Using IRR's: \textquotedblleft Choose the Project
with the Highest IRR\textquotedblright}

When considering the choice of projects one must be explicit about the
interrelationships between the projects. Is it a situation where one can
choose several projects out of a set of projects (i.e., choose all projects
with positive $NPV$) or is one restricted to choosing only one project out of
the set (i.e., choose the project with highest $NPV$). The alleged rule
\textquotedblleft Choose the project with the highest IRR\textquotedblright%
\ is usually applied in the situation where one can only choose one project
out of the set of alternatives (e.g., build only one building on a site).

It is easy to see the fallacy if the projects are of quite different scale.
Suppose one project turns $\$100$ into $\$200$ in one year for an $IRR$ of
$100\%$ while another project turns $\$1000$ into $\$1500$ in a year for an
IRR of only $50\%$. If one must choose one project or the other (and cannot
repeat the first project ten times), then clearly the second project is more
profitable (assuming a discount rate less than $50\%$) even though it has the
lower $IRR$.

To be taken seriously, the \textquotedblleft Highest $IRR$\textquotedblright%
\ rule might be amended to read: \textquotedblleft Among projects with the
same required investment capital, choose the project with the highest
$IRR$.\textquotedblright\ This amended rule is also wrong as can be seen by
comparing projects A and B in the previous table. Both have the same invested
capital of $\$1000$ and project B has the higher $IRR$ ($23.38\%$ versus
$20\%$). But at the discount rate of $10\%$ (or lower rates), project A has
the higher $NPV$ so it is the best project at those discount rates.

Perhaps the "Highest $IRR$" rule seems attractive because many practitioners
incorrectly extrapolate the rule from the case of one-year projects (only one
cash payout) to multi-year projects. The Highest $IRR$ rule works for projects
with the same initial capital investment and only one cash payout at the end
of the period. Then it is, of course, true that the project with the highest
cash payout is the best project (although both projects might have negative
$NPV$ at high discount rates).

If there is a multi-year payout, then projects begin to differ in more subtle
ways. Some projects pay out early while others pay out later but in greater
amounts. To know which is best, one must know how heavily to discount the
future payouts--which means one must use the discount rate to compute the
$NPV$. Thus it is easy to see that the multi-year highest $IRR$ rule could not
possibly be valid since it makes no mention of the discount rate.

\subsection{Pitfall 3 in Using IRR's: Multiple IRR's}

It is unfortunately possible for a project to have two or more $IRR$'s.
However, this can only happen if the cashflows changes signs more than once
(e.g., go from negative to positive and then back to negative). Then the $NPV$
curve could cross the horizontal axis twice giving two $IRR$'s.

\begin{center}%
\begin{tabular}
[c]{|c|c|c|c|c|c|c|c|}\hline
Project & $C_{0}$ & $C_{1}$ & $C_{2}$ & $C_{3}$ & $IRR1$ & $IRR2$ & $NPV$ @
$30\%$\\\hline\hline
D & $-1000$ & $1450$ & $1500$ & $-2200$ & $28.52\%$ & $39.34\%$ &
$\$1.59$\\\hline
\end{tabular}

Table 10: Project D has multiple $IRR$'s.
\end{center}

\begin{figure}[h]
	\centering
	\includegraphics[width=0.7\linewidth]{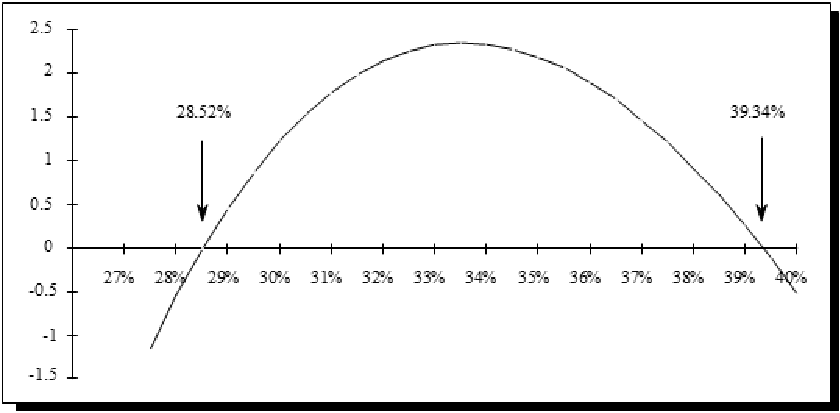}
	\caption{Project with multiple $IRR$'s}
	\label{fig:multi-irrs}
\end{figure}

Project D starts out with an investment of $\$1000$ in Table 10, has two
positive cash payouts, and then has a large negative closing cost of $\$2200$
(e.g., cleaning up the environment after a project is finished). The multiple
$IRR$'s are illustrated in Figure \ref{fig:multi-irrs}.

The project has two $IRR$'s at about $28.52\%$ and $39.34\%$. In between, the
project has a small positive $NPV$.

It might be noted that a project might have no $IRR$ instead of multiple
$IRR$'s. For instance, if we lower the payout $C_{2}$ in project D from $1500$
to $1450$, then the $NPV$ curve shifts down enough that it does not cross the
horizontal axis at all so it has no $IRR$.

\subsection{Criterion for Pair-wise Choice Between Projects}

When can $IRR$'s be used to make choices between investment projects? Under
certain assumptions, the $IRR$ concept can be used to make a choice between
two mutually exclusive projects. We will assume that for both projects, the
$NPV$'s decline as the discount rate increases.

Suppose we are given a choice between two projects such as projects A and B
previously considered. Table 11 gives the calculation for the project A-B.

\begin{center}%
\begin{tabular}
[c]{|c|c|c|c|c|c|c|c|}\hline
Project & $C_{0}$ & $C_{1}$ & $C_{2}$ & $C_{3}$ & $IRR$ & $NPV$ @ $10\%$ &
$NPV$ @ $12\%$\\\hline\hline
A & $-1000$ & $200$ & $200$ & $1200$ & $20\%$ & $\$248.69$ & $\$192.15$%
\\\hline
B & $-1000$ & $500$ & $500$ & $500$ & $23.38\%$ & $\$243.43$ & $\$200.92$%
\\\hline
A-B & $0$ & $-300$ & $-300$ & $700$ & $10.73\%$ & $\$5.26$ & $-\$8.77$\\\hline
\end{tabular}

Table 11: Pair-wise comparison of Projects A and B.
\end{center}

The decision, of course, will depend on the discount rates. At $10\%$, A is
the best project--while at $12\%$, B is the best project. What is the cutoff
interest rate at which one project is replaced by the other as the best
project? The cutoff interest rate is found by considering the IRR of the
\textquotedblleft difference project\textquotedblright\ A-B. The $IRR$ of A-B
is about $10.73\%$ which means that for interest rates below that (such as
$10\%$), project A is best, while for interest above that rate (such as
$12\%$), project B is best.

One might ask, why choose A--B as the difference project? Why not B-A? The
answer is that the difference project should also satisfy our rule that the
$NPV$ declines as the discount rate increases. Project A-B satisfies the rule
while B-A does not. This can be seen from the pattern of the signs in the
cashflows. If the cashflows go from negative to positive as time increases,
and do not reverse later on, then the $NPV$ curve will slope downward. Since
the difference project A-B has that property, we say the \textquotedblleft A
is later than B\textquotedblright\ in the sense that A's payouts are
unambiguously later than the payouts from B.

Since A is later than B, it can be intuitively understood why A is better
before--and B after, the cutoff point of $10.73\%$. As the discount rate
increases above $10.73\%$, both projects lose $NPV$ (downward sloping $NVP$
curve) but A loses $NPV$ faster since its payouts are later and will thus be
hit harder by the higher discount rate. The reverse happens as the discount
rate decreases below the cutoff point.

It is also possible to understand the pair-wise choice rule in terms of our
previous result that a project (with downward sloping $NPV$) is profitable if
its $IRR$ exceeds the discount rate. The difference project A--B can be
thought of as the project of converting from B to A. If the discount rate is
below the cutoff point of $10.73\%$, which is the $IRR$ of the difference
project (with downward sloping $NPV$), then it is profitable to convert from B
to A, i.e., A is better than B. If the discount rate exceeds the cutoff point,
then it is unprofitable to convert from B to A, i.e., B is better than A (for
more analysis, see Chapter 3 \cite{hirshleifer:capital}).

A final word of warning about what defines or delimits a \textquotedblleft
project.\textquotedblright\ In particular, different initially defined
projects could have virtually the same investment and cash returns, but they
could offer quite different further investment opportunities for the cash
throwoffs. Then a broader conception of the projects to include the
reinvestment of the cash returns could reveal that one of the reconceived
projects was clearly superior to the other \cite{kelleher:irr}.

\section{Appendix 1: Proof of the General Linear Recurrence Formula}

Consider the general linear recurrence relation defined by

\begin{center}
$y_{0}=c$ and $y_{k}=my_{k-1}+b$ for some constants $m$, $b$, and $c$.
\end{center}

The general solution has the form

\begin{center}
$y_{n}=m^{n}c+m^{n-1}b+...+mb+b$
\end{center}

which can be expressed by the formula

\begin{center}
$y_{n}=\left\{
\begin{array}
[c]{c}%
m^{n}c+\frac{b\left(  m^{n}-1\right)  }{m-1}\text{ for }m\neq1\\
c+nb\text{ for }m=1
\end{array}
.\right.  $
\end{center}

Taking the $k^{th}$ year's income as $y_{k}$ for $k=1,...,n$, the present
value of the income stream is:

\begin{center}
$V_{n}=\sum_{k=1}^{n}\frac{y_{k}}{\left(  1+i\right)  ^{k}}$.
\end{center}

A formula for this summation will be derived for each of the four cases where
$m$ equals or does not equal $1$ and $1+i$.

\subsection{Case 1: $m\neq1,1+i$}

Expanding the summation yields:

\begin{center}
$V_{n}=\sum_{k=1}^{n}\frac{y_{k}}{\left(  1+i\right)  ^{k}}=\sum_{,=1}%
^{n}\frac{m^{k}c+m^{k-1}b+...+mb+b}{\left(  1+i\right)  ^{k}}$

$=c\sum_{k=1}^{n}\left(  \frac{m}{1+i}\right)  ^{k}+b\sum_{k=1}^{n}%
\frac{\left(  m^{k}-1\right)  /\left(  m-1\right)  }{\left(  1+i\right)  ^{k}%
}$

$=\left[  \frac{b}{m-1}+c\right]  \sum_{k=1}^{n}\left(  \frac{m}{1+i}\right)
^{k}-\frac{b}{m-1}a_{n}$.
\end{center}

Since $m\neq1+i$, the summation in the last term can be simplified.

\begin{center}
$V_{n}=\left[  \frac{b}{m-1}+c\right]  \frac{m\left[  1-\left(  \frac{m}%
{1+i}\right)  ^{n}\right]  }{1+i-m}-\frac{b}{m-1}a_{n}$.

Case 1 Formula
\end{center}

\subsection{Case 2: m = 1+i $\neq$ 1}

In this case we can easily evaluate the summation:

\begin{center}
$\sum_{k=1}^{n}\left(  \frac{m}{1+i}\right)  =n$
\end{center}

and $m-1=i$, so the last step of the Case $1$ derivation can be easily
modified to yield the desired formula.

\begin{center}
$V_{n}=nc+\frac{b\left[  n-a_{n}\right]  }{i}$.

Case 2 Formula
\end{center}

There is some other useful information that can be extracted in this case and
that will be useful later. Since $m=1+i$, we have that $y_{k}=(1+i)^{k}%
c+s(k,i)b=(1+i)^{k}c+s_{k}b$ so the value $V_{n}$ can be expressed as follows:

\begin{center}
. $V_{n}=\sum_{k=1}^{n}\frac{(1+i)^{k}c+s_{k}b}{\left(  1+i\right)  ^{k}%
}=nc+b\sum_{k=1}^{n}\frac{s_{k}}{\left(  1+i\right)  ^{k}}=nc+b\sum_{k=1}%
^{n}a_{k}$.
\end{center}

From the case 2 formula we can thus derive the following:

\begin{center}
$\sum_{k=1}^{n}\frac{s_{k}}{\left(  1+i\right)  ^{k}}=\sum_{k=1}^{n}%
a_{k}=\frac{n-a_{n}}{i}$.
\end{center}

There is an interesting direct and intuitive proof of this formula using the
perpetuity capitalization formula. If there is the constant amount $n-a(n,i)$
at the end of each year in perpetuity, then the present value is the
right-hand side term: $[n-a(n,i)]/i$.

\begin{figure}[h]
\centering
\includegraphics[width=0.7\linewidth]{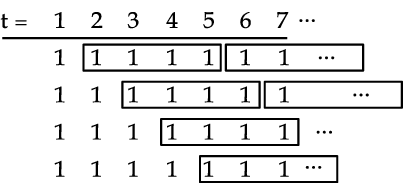}  \caption{Case 2
table}%
\label{fig:app1-table1s}%
\end{figure}

The picture below illustrates this proof for the case of $n=4$. There is an
array of four $1$'s at $t=1,2,...$ in perpetuity. Consider the column of four
$1$'s at $t=1$ and the top box of four $1$'s that begins at $t=2$. The value
of that box of $1$'s at $t=1$ is $a_{4}=a(4,i)$ and the value of the four
$1$'s in the column at $t=1$ is, of course, $4$. Thus the value of those $1$'s
minus the box is $4-a_{4}$ at $t=1$. Then consider the next column of four
$1$'s at $t=2$ and the second box of $1$'s that begins in the second row at
$t=3$. The value of those $1$'s minus that box at $t=2$ is again $4-a_{4}$ as
shown in Figure \ref{fig:app1-table1s}.

We continue in a similar way with the process cycling at $t=5$. The four $1$'s
in the column at $t=5$ are coupled with the second box in the top row starting
at $t=6$. The value of those $1$'s minus that box is $4-a_{4}$ at $t=5$. Since
this pattern repeats itself forever, the present value is $[4-a_{4}]/i$. But
all the $1$'s in boxes occurred both positively (in their column) and
negatively (in their box) so they cancel out. Thus only the $1$'s not in any
box contribute to the total value, and their present value is clearly
$a_{1}+a_{2}+a_{3}+a_{4}$. Thus we have shown that:

\begin{center}
$\sum_{k=1}^{4}a\left(  k,i\right)  =\frac{[4-a_{4}]}{i}$.
\end{center}

Although illustrated for the $n=4$ case, the pattern of the proof clearly
works for any $n$.

\subsection{Case 3: m = 1, m $\neq$ 1+i}

In this case, $y_{k}=c+nb$ so the summation for $V_{n}$ yields:

\begin{center}
$V_{n}=ca_{n}+b\sum_{k=1}^{n}\frac{k}{\left(  1+i\right)  ^{k}}$
\end{center}

\begin{figure}[h]
\centering
\includegraphics[width=0.3\linewidth]{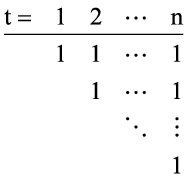}  \caption{Case 3
Table 1}%
\label{fig:case3-table1s}%
\end{figure}

In the summation of the terms $k/(1+i)^{k}$ each such term is the present
value of a $1$ in a column of the following triangular array ($n$ rows and $n$
columns) as illustrated in Figure \ref{fig:case3-table1s}.

Summing the present values by rows, we have:

\begin{center}
$\sum_{k=1}^{n}\frac{k}{\left(  1+i\right)  ^{k}}=a_{n}+\frac{a_{n-1}}{\left(
1+i\right)  ^{1}}+...+\frac{a_{1}}{\left(  1+i\right)  ^{n-1}}$

$=\frac{1-1/\left(  1+i\right)  ^{n}}{i}+\frac{1-1/\left(  1+i\right)  ^{n-1}%
}{i\left(  1+i\right)  ^{1}}+...+\frac{1-1/\left(  1+i\right)  ^{1}}{i\left(
1+i\right)  ^{n-1}}$

$=\frac{\left(  1+i\right)  a_{n}}{i}-\frac{n/\left(  1+i\right)  ^{n}}{i}$
\end{center}

Adding and subtracting $n/i$ allows us to simplify the sum to:

\begin{center}
$\sum_{k=1}^{n}\frac{k}{\left(  1+i\right)  ^{k}}=\frac{\left(  1+i\right)
a_{n}}{i}-\frac{n/\left(  1+i\right)  ^{n}}{i}+\frac{n}{i}-\frac{n}{i}=\left(
n+1\right)  a_{n}-\frac{\left[  n-a_{n}\right]  }{i}$.
\end{center}

\begin{figure}[h]
\centering
\includegraphics[width=0.5\linewidth]{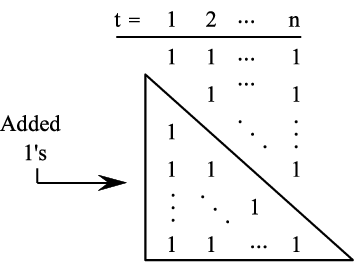}  \caption{Case 3 Table
2}%
\label{fig:case3-table2}%
\end{figure}

There is another way to arrive at this result. Suppose we complete the
triangular array used above by continuing $1$'s down each column to form an
$n\times n$ array and then add one more row of $1$'s at the bottom to form an
$(n+1)\times n$ array as seen in Figure \ref{fig:case3-table2}.

There are then $n+1$ rows each with the present value $a_{n}$. But we must
subtract the added $1$'s which have the present value $a_{1}+a_{2}%
+...+a_{n}=[na_{n}]/i$. Thus the original triangular array has the value of
the difference:

\begin{center}
$\sum_{k=1}^{n}\frac{k}{\left(  1+i\right)  ^{k}}=\left(  n+1\right)
a_{n}-\frac{n-a_{n}}{i}\qquad$
\end{center}

Substituting back into the formula for $V_{n}$ and rearranging finishes case
$3$.

\begin{center}
$V_{n}=\left[  c+\left(  n+1\right)  b\right]  a_{n}-\frac{b\left[
n-a_{n}\right]  }{i}$.

Case 3 Formula
\end{center}

\subsection{Case 4: m = 1 = 1+i}

Since$m=1$ and $i=0$, the original summation can be quickly simplified.

\begin{center}
$V_{n}=\sum_{k=1}^{n}\left(  c+kb\right)  =nc+\sum_{k=1}^{n}k$
\end{center}

The summation $1+2+...+n$ is easily evaluated by adding it to itself written
backwards as shown in Figure \ref{fig:euler-trick}:

\begin{figure}[h]
\centering
\includegraphics[width=0.4\linewidth]{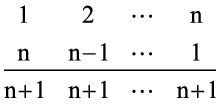}  \caption{Euler's
addition trick}%
\label{fig:euler-trick}%
\end{figure}

\noindent so the original sum is one-half that amount. Hence we arrive at the
formula for the last case.

\begin{center}
$V_{n}=nc+\frac{bn\left(  n+1\right)  }{2}$.

Case 4 Formula
\end{center}

\section{Appendix 2: Proof of the Main Theorem on Amortization Tables}

To prove the result,

\begin{center}
$\sum_{k=1}^{n}\frac{I_{k}}{\left(  1+i\right)  ^{k}}=\sum_{k=1}^{n}P_{k}$
\end{center}

\noindent where $I_{k}=P_{k}+i(P_{k}+...+P_{n})=(1+i)P_{k}+i(P_{k+1}%
+...+P_{n})$, we need to evaluate the sum:

\begin{center}
$\sum_{k=1}^{n}\frac{I_{k}}{\left(  1+i\right)  ^{,}}=\sum_{k=1}^{n}%
\frac{(1+i)P_{k}+i(P_{k+1}+...+P_{n})}{\left(  1+i\right)  ^{k}}$.
\end{center}

\begin{figure}[h]
\centering
\includegraphics[width=0.5\linewidth]{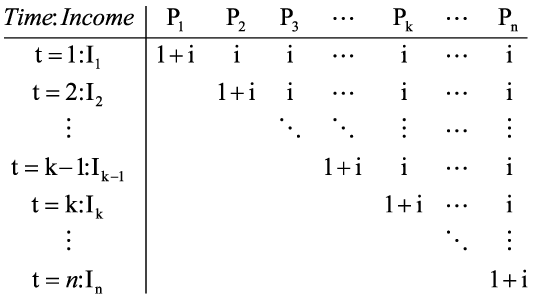}  \caption{Matrix for
proof of Main Theorem}%
\label{fig:app2-table1}%
\end{figure}

To rearrange the sum, we consider the following table of the terms to be
discounted at $t=1,2,...,n$. Each row gives the income for that time period,
the sum of the table entries across the row times the $P_{j}$'s at the head of
the columns as illustrated in Figure \ref{fig:app2-table1}.

We can now easily rewrite the sum as the discounted present value of the
entries in the columns to obtain:

\begin{center}
$\sum_{k=1}^{n}\frac{I_{k}}{\left(  1+i\right)  ^{,}}=\sum_{k=1}^{n}%
P_{k}\left(  \frac{1}{\left(  1+i\right)  ^{k}}+i\left[  \frac{1}{\left(
1+i\right)  ^{k}}+...+\frac{1}{\left(  1+i\right)  ^{1}}\right]  \right)  $

$=\sum_{k=1}^{n}P_{k}\left(  \frac{1}{\left(  1+i\right)  ^{k}}+ia_{k}\right)
=\sum_{k=1}^{n}P_{i}\left(  \frac{1}{\left(  1+i\right)  ^{k}}+1-\frac
{1}{\left(  1+i\right)  ^{k}}\right)  =\sum_{k=1}^{n}P_{k}$
\end{center}

\noindent which completes the proof of the Main Theorem on Amortization Tables.

\section{Statements}

There are no conflicts of interest or funding sources to report.


\section{Bibliography}

\begin{thebibliography}{9}                                                                                                %
\bibitem {friedman:reapp}Friedman, Jack P. and Nicholas Ordway 1988.
\textit{Income Property Appraisal and Analysis}. Englewood Cliffs: Prentice Hall.

\bibitem {hirshleifer:capital}Hirshleifer, Jack. 1970. \textit{Investment,
Interest, and Capital}. Englewood Cliffs: Prentice-Hall.

\bibitem {kelleher:irr}Kelleher, John, and Justin MacCormack. 2004.
\textquotedblleft Internal Rate of Return: A Cautionary
Tale.\textquotedblright\ \textit{McKinsey Quarterly}, August.
\end{thebibliography}
\end{document}